\documentclass[aps,showpacs]{revtex4}
\usepackage{graphicx}  
\usepackage{dcolumn}   
\usepackage{bm}        
\usepackage{amssymb}   
\usepackage{amsmath}
\usepackage{dsfont}

\newcommand{\tr}{{\textrm{Tr}\,}}
\newcommand{\be}{\begin{eqnarray}}
\newcommand{\ee}{\end{eqnarray}}
\newcommand{\pp}[1]{\phantom{#1}}
\newcommand{\uu}[1]{\underline{#1}}
\begin{document}

\title{Theory versus experiment for vacuum Rabi oscillations in lossy cavities (II):\\
Direct test of uniqueness of vacuum}
\author{Marcin Wilczewski and Marek Czachor}
\affiliation{
Katedra Fizyki Teoretycznej i Informatyki Kwantowej\\
Politechnika Gda\'nska, 80-952 Gda\'nsk, Poland,\\
Krajowe Centrum Informatyki Kwantowej, 81-824 Sopot, Poland\\
and\\
Centrum Leo Apostel (CLEA)\\
Vrije Universiteit Brussel, 1050 Brussels, Belgium
}

\begin{abstract}
The paper continues the analysis of vacuum Rabi oscillations we started in Part I [Phys. Rev. A {\bf 79}, 033836 (2009)].
Here we concentrate on experimental consequences for cavity QED of two different classes of representations of harmonic oscillator Lie algebras. The zero-temperature master equation, derived in Part I for irreducible representations of the algebra, is reformulated in a reducible representation that models electromagnetic fields by a gas of harmonic oscillator wave packets.  The representation is known to introduce automatic regularizations that in irreducible representations would have to be justified by ad hoc arguments. Predictions based on this representation are characterized in thermodynamic limit by a single parameter $\varsigma$, responsible for collapses and revivals of Rabi oscillations in exact vacuum. Collapses and revivals disappear in the limit $\varsigma\to\infty$. Observation of a finite $\varsigma$ would mean that cavity quantum fields are described by a non-Wightmanian theory, where vacuum states are zero-temperature Bose-Einstein condensates of a $N$-particle bosonic oscillator gas and, thus, are non-unique. The data collected in the experiment of Brune {\it et al.\/} [Phys. Rev. Lett. {\bf{76}}, 1800 (1996)] are consistent with any $\varsigma>400$.
\end{abstract}

\maketitle

\section{Introduction}

The first part of this work \cite{pra} was devoted to the problem of understanding the structure and origin of decoherences that occur if a single atom propagates through an initially empty optical cavity. We compared theoretical predictions with the most precise available data \cite{Brune}, and showed that some of the controversial issues discussed earlier in the literature can be resolved if one correctly takes into account the Gaussian structure of the cavity mode and the fact that the cavity was open.

As opposed to standard cavity QED approaches, modeling energy losses by jumps between bare energy eigenstates \cite{Carmichael,C-T}, we based our analysis on Markovian master equations whose Davies operators describe jumps between dressed states. Our approach is consistent with the general formalism of Davies \cite{Davies,Davies2}, or the more recent works of Scala {\it et al.\/} \cite{Scala,Scala2,Scala3}, where losses due to interaction with environment lead to transitions between eigenstates of the system Hamiltonian (guaranteeing stationarity of the asymptotic state at $T>0$ K). What is important, the choice of dressed states simplifies computations and naturally incorporates long-wavelength transitions within a single dressed-state manifold, an effect expected in open cavities. In order to derive the model from a microscopic level, we assumed the system-reservoir interaction  of the form $\big(\alpha (a+a^\dag)+\beta a^\dag a\big)\otimes B$, where $\alpha$, $\beta$ are parameters.

The part proportional to $\beta$ (let us refer to it as the Alicki interaction term \cite{Alicki}) is responsible for transitions within the same dressed-state manifold. It simultaneously makes the coupling between the system and the reservoir more sensitive to the photon number. Perhaps, it is the latter property of the interaction that leads to apparent underestimates of the cavity quality factor if $Q$ is measured with relatively strong fields, while the actual measurements of Rabi oscillations are performed in almost exact vacuum. The problem of $Q$ is one of the issues that require further experimental and theoretical studies.

Still, the list of open questions is longer, and some of them touch the very fundamentals of quantum field theory. For example, it is known that different physical systems in general correspond to different representations of Lie algebras. Fields are quantized by means of harmonic-oscillator Lie algebras, but can the cavity QED data tell us something about their representations? It turns out that there exists a class of physically motivated representations whose predictions are characterized, in certain thermodynamic limit, by a single parameter $\varsigma$ which influences vacuum Rabi oscillations.

Determination of $\varsigma$ is, in principle, within the reach of cavity QED experiments. The first estimates on $\varsigma$, $\varsigma>200$, were given in \cite{theorphys}, but the approach to decoherence was not in that paper based on systematically derived master equations. As such, it was not reliable, a fact that motivated the research project whose partial results were reported in \cite{pra}, and now completed in the present paper. Basing the analysis of decoherence on the results from \cite{pra}, we will show that the data from the experiment of Brune {\it et al.\/} \cite{Brune} are consistent with any $\varsigma>400$. We will also show that a finite value of $\varsigma$ implies collapses and revivals of Rabi oscillation even in exact vacuum. The first revival time is
$
t_r
=
\big( \varsigma+ \sqrt{ \varsigma(\varsigma-1)}\big)T_{\rm Rabi}
$,
where $T_{\rm Rabi}$ is the period of Rabi oscillation.

Observation of the revival in exact vacuum could be a proof of a non-Wightmanian nature of cavity QED. Conceptual consequences of such a finding might be enormous but --- paradoxically ---  implications for agreement between standard theory and experiment could be smaller from what one might expect at a first glance. The reason is the correspondence principle stating that the weak law of large numbers, $N\to\infty$ with $Z=$~const, maps theories based on our reducible representations into regularized forms of those based on irreducible representations. The limiting forms are already regularized, that is, the automatic cut-offs occur in exactly those places where in standard approaches one puts them by hand. This is a strong argument in favor of field quantization in terms of reducible representations of harmonic oscillator Lie algebras.

The paper is organized as follows. In Sec.~II we discuss the physical background of the reducible representation in question. In Sec.~III we show how to decompose the reducible representation into blocks that allow us to perform calculations, in each block separately, by means of the methods known from the standard formalism. In Sec.~IV we describe in detail the block structure of dressed states. We show that the number of dressed states in a given block is $s+2$, where $s$ is a parameter that characterizes the block. In Sec.~V we generalize to the reducible representation the derivation of an appropriate master equation; the strategy is the same as in \cite{pra}, only the representation of the Lie algebra is different. As opposed to \cite{pra}, we concentrate on zero-temperature master equation. Therefore, we had to supplement this section be a technical Appendix on $T=0$ irreducible-representation solutions (given at the end of the paper). An extension to $T>0$ would be immediate, but it would simultaneously introduce a number of irrelevant technical details, thus obscuring the main message of the paper. In Sec.~VI we discuss energy losses and compare the reducible treatment with the one based on general irreducible representations. The goal of this section is to give an alternative proof of necessity of renormalizing the decay parameters occurring in the master equations, before one compares theory with experiment. Sec.~VII contains the main result of the paper: experiments can, in principle, discriminate between irreducible and reducible representations, and we show how to estimate the relevant experimental parameters.

\section{Preliminaries}

In order to understand the wider context of the issue, let us return to the Hamiltonian we employed in \cite{pra},
\be
H
&=&
\hbar
\Big(
\frac{\omega}{2}\sigma_3
+
\omega a^\dag a
+
{q}(\sigma_-a^\dag+\sigma_+a)
\Big)
\nonumber\\
&\pp=&+
\big(\alpha (a+a^\dag)+\beta a^\dag a\big)B
+
H_R.\label{H1}
\ee
$\sigma_3$, $\sigma_\pm=(\sigma_1\pm i\sigma_2)/2$ are the Pauli matrices, $a$ is the usual harmonic-oscillator annihilation operator, the coupling parameter ${q}$ is, for simplicity, assumed to be real, and $B$ and $H_R$ are operators whose explicit form is irrelevant since they correspond to the reservoir. It is known, at least since the work of Tavis and Cummings \cite{TC}, that (\ref{H1}) is the simplest case of a more abstract Hamiltonian
\be
H
&=&
\hbar
\Big(
\omega J_3
+
\omega a^\dag a
+
{q}(J_-a^\dag+J_+a)
\Big)
\nonumber\\
&\pp=&+
\big(\alpha (a+a^\dag)+\beta a^\dag a\big) B
+
H_R,\label{HN}
\ee
where $J_3$, $J_\pm$ are elements of the Lie algebra su(2),
\be
{[J_3,J_\pm]} &=& \pm J_\pm,\\
{[J_-,J_+]} &=& 2 J_3.
\ee
Replacing a single two-level atom by a system of several two-level atoms, one finds that $J_3$, $J_\pm$ are given by a higher-spin, reducible representation of su(2).

From the point of view of a Lie-algebraic purist, the Hamiltonian (\ref{HN}) mixes abstract elements $J_3$, $J_\pm$ of su(2) with a concrete {\it representation\/} of another Lie algebra. Indeed, the operators $a_-=a$, $a_+=a^\dag$, $a_3=a^\dag a$, and $a_0=1$ are a representation of the one-dimensional harmonic-oscillator Lie algebra
\cite{ho}, ho(1),
\be
{[a_-,a_+]} &=& a_0,\\
{[a_\pm,a_0]} &=& [a_3,a_0]=0,\\
{[a_3,a_\pm]} &=& \pm a_\pm.
\ee
The abstract Lie-algebraic generalization of (\ref{H1}) is thus \cite{ho'}
\be
H
&=&
\hbar
\Big(
\omega J_3
+
\omega a_3
+
{q}(J_-a_+ +J_+a_-)
\Big)
\nonumber\\
&\pp=&+
\big(\alpha (a_-+a_+)+\beta a_3\big) B
+
H_R,.\label{H-Lie}
\ee
The simplest example of a representation of (\ref{H-Lie}) is a single spin-1/2 particle interacting with a single one-dimensional harmonic oscillator (both su(2) and ho(1) are then given by irreducible representations). The Tavis-Cummings system is formally equivalent to several spin-1/2 systems interacting with a single harmonic oscillator (su(2) is then given by a reducible representation incorporating different spins, but ho(1) is still represented irreducibly). The atom-field system we consider in the present paper is dual to the Tavis-Cummings model: its formal equivalent is a single spin-1/2 system interacting with several harmonic-oscillator {\it wave packets\/}.

The reason for reducibility  of the $N$-atom representation of su(2) discussed in \cite{TC} is mathematically very deep: multi-atomic Hilbert spaces are described by tensor products of single-atom representations, but tensor products of irreducible representations of su(2) are reducible. So, physically, it is the multi-particle structure of the $N$-atom system that forces the representation of su(2) to be reducible.

Now, for a quantum optician a cavity quantum field is an ensemble of {\it many\/} harmonic oscillators.  The relevant algebraic structure is given by the multi-particle harmonic oscillator Lie-algebra ho$(m)$, with $m=\infty$ not excluded,
\be
{[a_-(\bm k),a_+(\bm k')]} &=& \delta_{\bm k,\bm k'}a_0(\bm k),\\
{[a_\pm(\bm k),a_0(\bm k')]} &=& [a_3(\bm k),a_0(\bm k')]=0,\\
{[a_3(\bm k),a_\pm(\bm k')]} &=& \pm \delta_{\bm k,\bm k'}a_\pm(\bm k).
\ee
The standard representation employed in quantum optics has the form typical of irreducible representations, with $a_0(\bm k)=1$.
But, in light of what we have written above, a multi-particle system should, in general, be represented reducibly. So, is $a_0(\bm k)=1$ really obvious? In what follows we will see that various regularizations that plague quantum field theoretic calculations may indicate something exactly opposite. In our opinion the issue is essential for a correct formulation of field quantization and is, in principle, testable in experiments with vacuum Rabi oscillations. Let us explain this viewpoint in more detail.

There is a simple argument showing that if the harmonic oscillators forming the field are described by wave packets, the resulting representation of the harmonic-oscillator algebra is reducible. In order to show it, let us return to the Alicki-type Hamiltonian $H_{\hat \omega}=a^\dag a \otimes \hbar\hat\omega$ (in fact, discussed already in \cite{czachor1}), where $\hat\omega=\sum_\omega \omega|\omega\rangle\langle\omega|$ is some operator. $H_{\hat \omega}$ acts in the Hilbert space of states
$|\psi\rangle=\sum_{n,\omega}\psi_{n,\omega}|{n,\omega}\rangle$, spanned by the eigenvectors $|{n,\omega}\rangle$,
\be
H_{\hat \omega}|{n,\omega}\rangle
&=&
\hbar\omega a^\dag a|{n,\omega}\rangle
=
n\hbar\omega|{n,\omega}\rangle.
\ee
Let us note that $H_{\hat \omega}$ acts effectively as the usual Hamiltonian $\hbar\omega a^\dag a$ typical of the harmonic oscillator with frequency $\omega$, but this is true only in the subspace spanned by $|{n,\omega}\rangle$, with fixed $\omega$. An important difference between $\hbar\omega a^\dag a$ and $H_{\hat \omega}$ can be seen if one computes the average
\be
\langle\psi|H_{\hat \omega}|\psi\rangle
&=&
\sum_{n,\omega}n\hbar\omega|\psi_{n,\omega}|^2.
\ee
If $|\psi\rangle$ is entangled, then different frequencies may be related to different numbers of excitations.
Although $H_{\hat \omega}$ describes a single harmonic oscillator, the average looks as if we considered an ensemble of oscillators with different frequencies. An interpretation of these facts is obvious: $|\psi\rangle$ is a single-oscillator {\it wave packet\/}, and $\omega$ is not a parameter but an eigenvalue (a quantum number). So the Alicki-type Hamiltonian describes the usual quantum harmonic oscillator, but with quantized $\omega$. Quantization of $\omega$ becomes more natural if one recalls that typical $\omega$s occurring in quantum harmonic oscillators are functions of {\it observables\/}
(a magnetic field, or center-of-mass position operators, say).

Now, let us rewrite $H_{\hat \omega}$ as follows
\be
H_{\hat \omega}
&=&
\sum_\omega \hbar\omega a^\dag a \otimes |\omega\rangle\langle\omega|\\
&=&
\sum_\omega \hbar\omega a_\omega^\dag a_\omega =\sum_\omega H_\omega\label{H_omega}
\ee
where
\be
a_\omega
&=&
a \otimes |\omega\rangle\langle\omega|,\label{a_omega}\\
H_\omega
&=&
\hbar\omega a_\omega^\dag a_\omega
\ee
The decomposition (\ref{H_omega}) together with the fact that the eigenvalues of $H_\omega$ are $n\hbar\omega$ shows that $H_{\hat \omega}$ has many (but not all) properties typical of a quantum-field Hamiltonian. The operators $a_-(\omega)=a_\omega$, $a_+(\omega)=a^\dag_\omega$, $a_3(\omega)=a_\omega^\dag a_\omega$, and $a_0(\omega)=1\otimes |\omega\rangle\langle\omega|$, form a representation of the harmonic-oscillator Lie algebra
\be
{[a_-(\omega),a_+(\omega')]} &=& \delta_{\omega,\omega'}a_0(\omega),\label{CCR1}\\
{[a_\pm(\omega),a_0(\omega')]} &=& [a_3(\omega),a_0(\omega')]=0,\label{CCR2}\\
{[a_3(\omega),a_\pm(\omega')]} &=& \pm \delta_{\omega,\omega'}a_\pm(\omega),\label{CCR3}
\ee
which is reducible since $a_0(\omega)$ is a projector and not a multiple of the identity.

We have shown that a single-oscillator wave packet leads to a reducible representation of ho$(m)$, where $m$ is the dimension of the space spanned by the eigenvectors of $\hat\omega$. Let us make one point very clear already here: In this representation, $m$ in ho$(m)$ is not the number of oscillators, but the number of eigenvalues of $\hat \omega$ in the one-oscillator wave packet. In the next Section we will show that a representation of ho$(\infty)$ corresponding to $N$ single-oscillator wave packets, each of the oscillators existing in quantum superpositions of all the possible frequencies, is a natural candidate for an algebra of electromagnetic field operators.

Before we launch on detailed calculations, let us briefly explain why $1<N<\infty$ reducible representations of the harmonic-oscillator Lie algebra may be precisely what is needed for a well defined quantum field theory. Let us consider a single-oscillator Hamiltonian, but now with the vacuum-energy term included, i.e.
\be
H_{\hat \omega}
&=&
\sum_\omega \frac{\hbar\omega}{2}(a_\omega^\dag a_\omega+a_\omega a_\omega^\dag)
\\
&=&
\sum_\omega \hbar\omega\big(a_\omega^\dag a_\omega+ \frac{1}{2}a_0(\omega)\big).
\ee
In such representations vacuum is represented by the entire Hilbert subspace of all the states that are annihilated by all $a_\omega$. If the oscillators are bosons, the role of vacuum is played by a zero-temperature Bose-Einstein condensate, i.e. any state of the form
\be
|\uu 0_\psi\rangle
&=&
|0_\psi\rangle
\otimes
\dots
\otimes
|0_\psi\rangle,\\
|0_\psi\rangle
&=&
\sum_{\omega}\psi_{0,\omega}|{0,\omega}\rangle,\\
a_\omega|0_\psi\rangle
&=& 0 \textrm{ for all $\omega$}.
\ee
A gas of $N$ such noninteracting wave packets has the Hamiltonian
\be
\uu H_{\hat \omega}
&=&\sum_{j=1}^N H_{\hat \omega}^{(j)}
\ee
where $H_{\hat \omega}^{(j)}$ is the Alicki-type Hamiltonian acting in the Hilbert space of a $j$th wave packet. The average energy of the Bose-Einstein condensate (i.e. the energy of {\it a\/} vacuum) is
\be
\langle\uu 0_\psi|\uu H_{\hat \omega} |\uu 0_\psi\rangle
&=&
\frac{N}{2}\sum_\omega \hbar\omega |\psi_{0,\omega}|^2
=
\frac{ZN}{2}\sum_\omega \hbar\omega \chi_{\omega},\nonumber\\
\label{ZN1}
\ee
where $Z=\max_\omega\{|\psi_{0,\omega}|^2\}$, and $\chi_{\omega}=|\psi_{0,\omega}|^2/Z$, $\lim_{\omega\to\infty}\chi_{\omega}=0$, is the natural ultraviolet cut-off function resulting from normalizability, $\sum_\omega |\psi_{0,\omega}|^2=1$, of the condensate wave function.

The cut-off appears only because the vacuum state is normalizable --- it is hard to imagine in quantum theory a reason more fundamental. However, there is a price for it: vacuum states are in such theories non-unique (as all Bose-Einstein condensates). Since the existence of a unique vacuum state is one of the Wightman axioms \cite{PCT,Haag}, quantum field theories based on this type of reducible representations of ho$(\infty)$ have to be non-Wightmanian. In this context it is worthy of mentioning that the non-Wightmanian aspects of the theory are not in conflict with Poincar\'e covariance and gauge invariance of electrodynamics, but concrete technical forms of these conditions are quite different from what we are accustomed to (see \cite{MCKW}).

In formula (\ref{ZN1}) we encounter the characteristic product of two parameters, $\varsigma=ZN$. This single parameter will be, effectively, the only free element in our reducible-representation treatment of Rabi oscillations. Let us note that the (thermodynamic) limit $N\to\infty$ with $\varsigma=$~const, implies shifting the cut-off to infinity, since $Z\to 0$ is equivalent to $Z_\omega\to 0$ for all $\omega$s, with $\sum_\omega Z_\omega=1$. We will later see that although the thermodynamic limit makes vacuum energy divergent, the limiting form of vacuum Rabi oscillation is well defined and carries information about the value of $\varsigma$.

\section{Reducible representation and its decomposition into irreducible components}

Let us assume that the electromagnetic field is a gas consisting of $N$ indefinite-frequency bosonic oscillators of the type described in the
Introduction. The basis in the $N$-oscillator Hilbert space ${\cal H}$ is given by tensor products
\be
|n_{\omega_1}\dots n_{\omega_N}\rangle
&=&
|n_1,\omega_1\rangle\dots |n_N,\omega_N\rangle
\ee
where the $\omega$s belong to the set of all the frequencies allowed by the cavity boundary conditions. The vacuum at zero temperature is assumed to be the pure state
\be
|\uu O\rangle
&=&
|O\rangle\dots|O\rangle=\sum_{\omega_1\dots\omega_N}O_{\omega_1}\dots O_{\omega_N}|0_{\omega_1}\dots 0_{\omega_N}\rangle,\\
|O\rangle
&=&
\sum_\omega O_\omega |0_\omega\rangle,\\
\sum_\omega |O_\omega|^2
&=& 1.
\ee
Taking $a_\omega$ in the form (\ref{a_omega}), we assume that the atom-light system interacting with the reservoir will be described by the $N$-wave-packet reducible representation of the Hamiltonian
(\ref{H-Lie}), where
\be
a_-
&=&
\frac{1}{\sqrt{N}}
\Big(
a_\omega \otimes {I} \otimes \ldots \otimes {I}
+
\ldots
+
{I} \otimes \ldots \otimes {I} \otimes a_\omega
\Big)=\uu a_\omega,\label{uu CCR1}
\\
a_+
&=&
\frac{1}{\sqrt{N}}
\Big(
a_\omega^\dagger \otimes {I} \otimes \ldots \otimes {I}
+
\ldots
+
{I} \otimes \ldots \otimes {I} \otimes a^\dagger_\omega
\Big)=\uu a_\omega^\dag,\label{uu CCR2}\\
a_3
&=&
a_\omega^\dag a_\omega \otimes {I} \otimes \ldots \otimes {I}
+
\ldots
+
{I} \otimes \ldots \otimes {I} \otimes a_\omega^\dag a_\omega=\uu N_\omega,\label{uu CCR3}\\
a_0
&=&
\frac{1}{N}
\Big(
\big(1\otimes|\omega\rangle\langle\omega|\big) \otimes {I} \otimes \ldots \otimes {I}
+
\dots +
{I} \otimes \ldots \otimes {I} \otimes \big(1\otimes|\omega\rangle\langle\omega|\big)
\Big)=\uu I_\omega,\label{uu CCR4}
\ee
is a reducible representation of the Lie algebra (\ref{CCR1})--(\ref{CCR3}), and $J_\pm$, $J_3$ are given, as in \cite{pra}, by the spin-1/2 representation of su(2). The physical meaning of this representation can be inferred from the form of $a_3$, which is the sum of number operators of $N$ independent harmonic-oscillator wave packets, and for $N=1$ it reduces to the single-oscillator wave-packet representation from Section II. The representation was introduced in the context of quantum optics in \cite{czachor1}. Preliminary results on its implications for Rabi oscillations can be found in \cite{theorphys}.

Let us denote by ${\cal H}_{\omega_1\dots\omega_N}$ the subspace spanned by $|n_{\omega_1}\dots n_{\omega_N}\rangle$. Obviously,
\begin{eqnarray}
{\cal{H}}
&=&
\bigoplus_{\omega_1,\ldots,\omega_N}
{\cal{H}}_{\omega_1\ldots\omega_N}.
\label{eq:Hspace-decomposition}
\end{eqnarray}
For any sequence of frequencies $\omega_j$, and for any $\omega$, one finds that the subspace ${\cal H}_{\omega_1\dots\omega_N}$ is invariant under the action of (\ref{uu CCR1})--(\ref{uu CCR4}). The central elements $a_0=\uu I_\omega$ satisfy
\be
\uu I_\omega
|n_{\omega_1}\dots n_{\omega_N}\rangle
&=&
\frac{s}{N}
|n_{\omega_1}\dots n_{\omega_N}\rangle,
\ee
where $s$ is the number of occurrences of $\omega$ in the sequence $\omega_1,\ldots,\omega_N$. As we can see, $\uu I_\omega$ is the operator of frequency of successes, known from quantum laws of large numbers \cite{LLN1,LLN2,LLN3,LLN4}.
Moreover, if $s=0$, i.e. when none of  $\omega_1,\ldots,\omega_N$ equals $\omega$, then
\be
\uu a_\omega
|n_{\omega_1}\dots n_{\omega_N}\rangle
&=&
\uu a_\omega^\dag
|n_{\omega_1}\dots n_{\omega_N}\rangle
=0.
\ee
Note that there exist nontrivial states annihilated by creation operators.
In consequence, for any $N+1$ different frequencies $\omega_1\dots \omega_{N+1}$, one finds
\be
\uu a_{\omega_1}^\dag\dots \uu a_{\omega_{N+1}}^\dag {\cal H}=
\uu a_{\omega_1}\uu a_{\omega_2}^\dag\dots \uu a_{\omega_{N+1}}^\dag {\cal H}=\dots=
\uu a_{\omega_1}\dots \uu a_{\omega_{N+1}} {\cal H}=
0.
\ee
The full system-reservoir Hilbert space ${\cal H}_{S+R}$ also can be split into a direct sum of subspaces invariant under the action of
\be
H
&=&
\hbar
\Big(
\frac{\omega}{2}\sigma_3
+
\omega \uu N_\omega
+
{q}(\sigma_-\uu a_\omega^\dag+\sigma_+\uu a_\omega)
\Big)
\nonumber\\
&\pp=&+
\big(\alpha (\uu a_\omega+\uu a_\omega^\dag)+\beta \uu N_\omega\big)B
+
H_R.\label{full H}
\ee
Indeed, let ${\cal{H}}_{A}$ denote the 2-dimensional Hilbert space of atomic states, spanned by the excited and ground states, $|e\rangle$ and $|g\rangle$. Then
\be
{\cal H}_{S+R}
&=&
\bigoplus_{\omega_1,\ldots,\omega_N}
{\cal{H}}_{A}\otimes
{\cal{H}}_{\omega_1\ldots\omega_N}\otimes {\cal H}_R
\ee
where
\be
H\,{\cal{H}}_{A}\otimes
{\cal{H}}_{\omega_1\ldots\omega_N}\otimes {\cal H}_R
&\subset&
{\cal{H}}_{A}\otimes
{\cal{H}}_{\omega_1\ldots\omega_N}\otimes {\cal H}_R.
\ee
The above observations, together with the known statistical properties of the operator of frequency of successes, supplemented by the fact that the right side of
\be
{[\uu a_\omega,\uu a_\omega^\dag]} &=& \uu I_\omega
\ee
reduces in ${\cal{H}}_{A}\otimes {\cal{H}}_{\omega_1\ldots\omega_N}\otimes {\cal H}_R$ to multiplication by an appropriate $s/N$, make calculations in our reducible representation as easy as those in the irreducible one.

It it is interesting that the dynamics in each of the subspaces ${\cal{H}}_{A}\otimes {\cal{H}}_{\omega_1\ldots\omega_N}\otimes {\cal H}_R$ involves at most $N$ different frequencies even if the number of frequencies admitted by boundary conditions is infinite. There are, of course, infinitely many such subspaces, corresponding to all the possible distributions of $\omega_j$ into $N$ locations.

In \cite{pra} we assumed the dynamics starting with the initial condition $|e,0\rangle \langle e,0|$. Its reducible analogue is
\begin{eqnarray}
\rho(0)
&=&
|e,\underline{O}\rangle \langle e,\underline{O}|
\nonumber
\\
&=&
|e\rangle \langle e|
\otimes
\sum_{\omega_1,\ldots,\omega_N}
\sum_{\omega_1',\ldots,\omega_N'}
O_{\omega_1} \ldots O_{\omega_N}
O^\ast_{\omega_1'} \ldots O^\ast_{\omega_N'}
|0_{\omega_1}\ldots 0_{\omega_N}\rangle \langle 0_{\omega'_1}\ldots 0_{\omega'_N}|.
\label{eq:160}
\end{eqnarray}
The indexing sequence in ${\cal{H}}_{\omega_1\ldots\omega_N}$ uniquely determines the parameter $s$. Assuming temperature $T=0$ and  (\ref{eq:160}) as the initial condition, the resulting master equation (derived later) will involve only a subspace of ${\cal{H}}_{\omega_1\ldots\omega_N}$, namely, the one spanned by
the following $s+2$ orthonormal vectors
\begin{subequations}

\begin{eqnarray}
|e,0_{\omega_1\ldots\omega_N}\rangle
&=&
|e\rangle |0_{\omega_1}\ldots 0_\omega \ldots 0_\omega \ldots 0_{\omega_N}\rangle
=|1\rangle
,
\\
|g,1^{(1)}_{\omega_1\ldots\omega_N} \rangle
&=&
|g\rangle |0_{\omega_1} \ldots 1_\omega \ldots 0_\omega \ldots 0_{\omega_N}\rangle=|2\rangle,
\\
&\vdots&
\nonumber
\\
|g,1^{(s)}_{\omega_1\ldots\omega_N}\rangle
&=&
|g\rangle |0_{\omega_1} \ldots 0_\omega \ldots 1_\omega \ldots 0_{\omega_N}\rangle=|s+1\rangle,
\\
|g,0_{\omega_1\ldots\omega_N} \rangle
&=&
|g\rangle |0_{\omega_1} \ldots 0_\omega \ldots 0_\omega \ldots 0_{\omega_N}\rangle=|s+2\rangle.
\end{eqnarray}
\label{eq:def-states}
\end{subequations}
Clearly, $|g,1^{(i)}_{\omega_1\ldots\omega_N} \rangle$, $i\in\{1,\ldots,s\}$ denotes the state of the atom-field system in which the atom is in the lower state  $|g\rangle$, and out of $s$ oscillators whose frequency is $\omega$, it is the $i$th oscillator which is excited to the first energy level.

\section{Dressed states in ${\cal{H}}_{A}\otimes {\cal{H}}_{\omega_1\ldots\omega_N}$}

Denote by $\Pi_{\omega_1\ldots\omega_N}$ the projector on ${\cal{H}}_A\otimes {\cal{H}}_{\omega_1\ldots\omega_N}$, and by $\pi_{\omega_1\ldots\omega_N}$ the one on the $(s+2)$-dimensional subspace spanned by (\ref{eq:def-states}). The total Hamiltonian consists of three terms: the Jaynes-Cummings atom-field part $H_{JC}=\hbar\Omega$, the reservoir Hamiltonian $H_R$, and the system-reservoir interaction.
The dressed states relevant for our problem are the eigenstates of
\be
\Omega_{\omega_1\ldots\omega_N}=\Omega \pi_{\omega_1\ldots\omega_N}.
\ee
It is instructive to write in the basis (\ref{eq:def-states}) the matrix $\Omega(s)_{kl}=\langle k|\Omega_{\omega_1\ldots\omega_N}|l\rangle$,
\begin{eqnarray}
\Omega(s)
&=&
\left(
\begin{array}{cccccc}
\frac{ \omega }{ 2 } & \frac{ {q} }{ \sqrt{N} } & \frac{ {q} }{ \sqrt{N} } & \dots & \frac{ {q} }{ \sqrt{N} } & 0\\
\frac{ {q} }{ \sqrt{N} } &  \frac{\omega }{ 2 }   & 0 & \dots  & 0 & 0 \\
\frac{ {q} }{ \sqrt{N} } & 0 &  \frac{ \omega }{ 2 } & 0 &  \dots & 0 \\
\vdots & \vdots & \vdots & \ddots & \vdots & \vdots\\
\frac{ {q} }{ \sqrt{N} } & 0 & 0 & \dots &   \frac{ \omega }{2 } & 0 \\
0 & 0 & 0 & 0 & \dots  & - \frac{\omega }{ 2 }
\end{array}
\right).
\label{eq:Homegas}
\end{eqnarray}
As we can see, $\Omega(s)$ does not explicitly depend on the concrete values of frequencies that index $\Omega_{\omega_1\ldots\omega_N}$, but only on the number $s$ of times the resonant frequency $\omega$ occurs in the indexing sequence $\omega_1\ldots\omega_N$.
The eigenvalues are
\begin{subequations}
\begin{eqnarray}
\Omega_\pm(s)
&=&
\frac{ \omega }{ 2 }
\pm
{q} \sqrt{ \frac{ s }{ N }},
\\
\Omega_{1}(s)
&=&
\Omega_{2}(s)=\dots=\Omega_{s-1}(s)=\frac{ \omega }{ 2 },
\\
\Omega_0(s)
&=&
-
\frac{ \omega }{ 2 },
\end{eqnarray}
\end{subequations}
with the corresponding orthonormal eigenvectors
\begin{subequations}
\begin{eqnarray}
|\Omega_0 (s) \rangle
=
\left(
\begin{array}{c}
0\\0\\0\\0\\ \vdots \\0 \\1
\end{array}
\right),\quad
|\Omega_1 (s) \rangle
=
\left(
\begin{array}{c}
0\\ \frac{1}{\sqrt{2}} \\ - \frac{1}{\sqrt{2}} \\0 \\\vdots \\ 0 \\0
\end{array}
\right),\quad
\dots,\quad
|\Omega_k (s) \rangle
=
\left(
\begin{array}{c}
0\\ \frac{1}{\sqrt{k(k+1)}} \\ \vdots \\ \frac{1}{\sqrt{k(k+1)}} \\ - \sqrt{ \frac{k}{k+1}  } \\ \vdots \\0
\end{array}
\right),
\ee
\be
|\Omega_{s-1} (s) \rangle
=
\left(
\begin{array}{c}
0\\ \frac{1}{\sqrt{(s-1)s}} \\ \frac{1}{\sqrt{(s-1)s}} \\ \vdots \\ \frac{1}{\sqrt{(s-1)s}} \\ - \sqrt{ \frac{s-1}{s}  } \\0
\end{array}
\right)
\\
\end{eqnarray}\label{eq:red:60}
and
\begin{eqnarray}
|\Omega_- (s) \rangle
=
\frac{1}{\sqrt{2s}}
\left(
\begin{array}{c}
- \sqrt{s} \\ 1\\\vdots \\1 \\0
\end{array}
\right),
\thickspace
|\Omega_+ (s) \rangle
=
\frac{1}{\sqrt{2s}}
\left(
\begin{array}{c}
\sqrt{s} \\ 1\\\vdots \\1 \\0
\end{array}
\right).
\end{eqnarray}\label{eq:red:61}
\end{subequations}
To avoid possible confusion let us remind that equations (\ref{eq:red:60}) define an orthonormal basis in the relevant subspace of ${\cal{H}}_{A}\otimes {\cal{H}}_{\omega_1\ldots\omega_N}$, and not in the full Hilbert space $\cal{H}$. The dressed states are related to the bare ones (\ref{eq:def-states}) by
\begin{subequations}
\begin{eqnarray}
|\Omega_0(s)\rangle
&=&
|g,0_{\omega_1\ldots\omega_N}\rangle,
\\
|\Omega_+(s)\rangle
&=&
\frac{1}{\sqrt{2}}
\Big(
\frac{1}{\sqrt{s}} \sum_{i=1}^{s} |g,1^{(i)}_{\omega_1\ldots\omega_N}\rangle
+
|e,0_{\omega_1\ldots\omega_N}\rangle
\Big),
\\
|\Omega_-(s)\rangle
&=&
\frac{1}{\sqrt{2}}
\Big(
\frac{1}{\sqrt{s}} \sum_{i=1}^s |g,1^{(i)}_{\omega_1\ldots\omega_N}\rangle
- |e,0_{\omega_1\ldots\omega_N}\rangle
\Big),
\\
|\Omega_{k}(s) \rangle
&=&
\frac{1}{ \sqrt{k(k+1)} }
\sum_{i=1}^k
|g,1^{(i)}_{\omega_1\ldots\omega_N}\rangle
-
\sqrt{\frac {k}{k+1} }
|g,1^{(k+1)}_{\omega_1\ldots\omega_N}\rangle;
\quad
k=1,\dots,s-1.
\end{eqnarray}
\label{eq:dressedstates}
\end{subequations}

\section{Master equation at zero temperature}

We are looking for a density matrix $\rho(t)$, acting in ${\cal H}_A\otimes {\cal H}$, satisfying the initial condition (\ref{eq:160}). Since our goal is to compute the atomic ground-state probability,
\be
p_g(t)
&=&
\tr |g\rangle\langle g|\rho(t)\nonumber\\
&=&
\sum_{\omega_1\dots\omega_N}
\sum_{\omega'_1\dots\omega'_N}
\tr |g\rangle\langle g|
\Pi_{\omega_1\dots\omega_N}
\rho(t)
\Pi_{\omega'_1\dots\omega'_N}\nonumber\\
&=&
\sum_{\omega_1\dots\omega_N}
\tr |g\rangle\langle g|
\Pi_{\omega_1\dots\omega_N}
\rho(t)
\Pi_{\omega_1\dots\omega_N}\nonumber\\
&=&
\sum_{\omega_1\dots\omega_N}
\tr |g\rangle\langle g|
\pi_{\omega_1\dots\omega_N}
\rho(t)
\pi_{\omega_1\dots\omega_N},
\ee
we have to derive and solve an effective master equation for the projected density matrix
\be
\rho_{\omega_1\dots\omega_N}(t)
&=&
\pi_{\omega_1\dots\omega_N}
\rho(t)
\pi_{\omega_1\dots\omega_N}.
\ee
The Jaynes-Cummings Hamiltonian can be analogously split into
\be
H_{JC}
&=&
\sum_{\omega_1\dots\omega_N}
\Pi_{\omega_1\dots\omega_N}H\\
&=&
\sum_{\omega_1\dots\omega_N}
H_{\omega_1\dots\omega_N}.
\ee
Each block $H_{\omega_1\dots\omega_N}$ can be yet further decomposed,
\be
H_{\omega_1\dots\omega_N}
&=&
H'_{\omega_1\dots\omega_N}
+
H''_{\omega_1\dots\omega_N}, \\
H'_{\omega_1\dots\omega_N}
&=&
\pi_{\omega_1\dots\omega_N}H_{\omega_1\dots\omega_N}.
\ee
Let ${\cal P}(\epsilon_{\omega_1\dots\omega_N})$ denote the spectral projectors of
\be
H_{\omega_1\dots\omega_N}=\hbar\sum_{\epsilon_{\omega_1\dots\omega_N}} \epsilon_{\omega_1\dots\omega_N} {\cal P}(\epsilon_{\omega_1\dots\omega_N}).
\ee
The spectrum of $H$ is inifinitely degenerated since each $H'_{\omega_1\dots\omega_N}$ has eigenvalues described in the previous section, and there are inifinitely many sequences $\omega_1\dots\omega_N$ corresponding to the same $s$. So let
\be
{\cal P}(\epsilon)
&=&
\sum_{\epsilon_{\omega_1\dots\omega_N}=\epsilon}{\cal P}(\epsilon_{\omega_1\dots\omega_N}).
\ee
Taking into account the system-reservoir interaction Hamiltonian
\be
AB=\big(\alpha (\uu a_\omega+\uu a_\omega^\dag)+\beta \uu N_\omega\big)B,
\ee
we find that the full density matrix $\rho(t)$ satisfies  at $T=0$~K the usual Markovian master equation \cite{Petruccione}
\begin{eqnarray}
\dot{\rho}
&=&
-i
\big[\Omega,\rho \big]
+
\sum_{\omega>0}
\gamma(\omega)
\left(
A(\omega) \rho A^\dagger(\omega)
-
\frac 1 2
\big[
A^\dagger(\omega) A(\omega), \rho
\big]_+
\right)
\label{eq:me-general0}
\end{eqnarray}
where
\be
A(\omega)
&=&
\sum_{\epsilon'-\epsilon=\omega}{\cal P}(\epsilon)A{\cal P}(\epsilon').
\ee
Since
\be
[\pi_{\omega_1\dots\omega_N},\Omega]=[\pi_{\omega_1\dots\omega_N},{\cal P}(\epsilon)]=
[\pi_{\omega_1\dots\omega_N},A]=0
\ee
we arrive at
\begin{eqnarray}
\dot{\rho}{}_{\omega_1\dots\omega_N}
&=&
-i
\big[\Omega_{\omega_1\dots\omega_N},\rho_{\omega_1\dots\omega_N} \big]
\nonumber\\
&+&
\sum_{\omega>0}
\gamma(\omega)
\left(
A_{\omega_1\dots\omega_N}(\omega) \rho_{\omega_1\dots\omega_N} A_{\omega_1\dots\omega_N}^\dagger(\omega)
-
\frac 1 2
\big[
A_{\omega_1\dots\omega_N}^\dagger(\omega) A_{\omega_1\dots\omega_N}(\omega), \rho_{\omega_1\dots\omega_N}
\big]_+
\right),\nonumber\\
\label{eq:me-general}
\end{eqnarray}
with
\begin{eqnarray}
A_{\omega_1\dots\omega_N}(\omega)
&=&
\sum_{\epsilon'_{\omega_1\dots\omega_N}-\epsilon_{\omega_1\dots\omega_N} =\omega}\pi_{\omega_1\dots\omega_N}
{\cal P}(\epsilon_{\omega_1\dots\omega_N}) A{\cal P}(\epsilon'_{\omega_1\dots\omega_N}).
\end{eqnarray}
There are no jumps between energy eigenstates belonging to subspaces indexed by different sequences $\omega_1\dots\omega_N$.
Employing (\ref{eq:dressedstates}) and the explicit form of $A$, we obtain
\begin{subequations}

\begin{eqnarray}
A_{\omega_1\dots\omega_N}( \Omega_+(s) - \Omega_0(s) )
&=&
\frac{ \alpha }{  \sqrt{2} }
\sqrt{ \frac s N }
|\Omega_0(s) \rangle \langle \Omega_+ (s)|,
\label{eq:51}
\\
A_{\omega_1\dots\omega_N}(\Omega_+(s) - \Omega_-(s) )
&=&
\frac{ \beta }{ 2 }
|\Omega_-(s) \rangle  \langle \Omega_+(s) |
\label{eq:52}
\\
A_{\omega_1\dots\omega_N}(\Omega_-(s)  - \Omega_0(s) )
&=&
\frac{ \alpha }{\sqrt{2}}
\sqrt{ \frac s N }
|\Omega_0(s) \rangle \langle \Omega_-(s) |,
\label{eq:53}
\\
A_{\omega_1\dots\omega_N}(\Omega_+(s) - \Omega_k(s) )
&=&
A_{\omega_1\dots\omega_N}(\Omega_k(s) - \Omega_-(s) )
\nonumber\\
&=&
A_{\omega_1\dots\omega_N}( \Omega_k(s)  - \Omega_0(s) )
=
0,
\quad
k=1,\ldots,s-1.
\label{eq:54}
\end{eqnarray}
\end{subequations}
An important and rather unexpected result is that jumps involving $|\Omega_k(s)\rangle$, $k=1,\ldots,s-1$, are not allowed \cite{no jumps}. This is why the master equation for $\rho_{\omega_1\dots\omega_N}(t)$ reads
\begin{eqnarray}
\dot{\rho}_{\omega_1\dots\omega_N}
&=&
-i [\Omega_{\omega_1\dots\omega_N},\rho_{\omega_1\dots\omega_N}]
\nonumber
\\
&+&
\gamma (\Omega_+(s) - \Omega_0(s)) \alpha^2
\frac s N
\nonumber\\
&\pp+&\times
\left(
\frac 1 2
|\Omega_0(s)\rangle \langle \Omega_+(s)| \rho_{\omega_1\dots\omega_N} |\Omega_+(s)\rangle \langle \Omega_0(s)|
-
\frac 1 4
[|\Omega_+(s)\rangle\langle \Omega_+(s)|,\rho_{\omega_1\dots\omega_N}]_+
\right)
\nonumber
\\
&+&
\gamma(\Omega_-(s) - \Omega_0(s))  \alpha^2
\frac s N
\nonumber\\
&\pp+&\times
\left(
\frac 1 2
|\Omega_0(s)\rangle \langle \Omega_-(s)| \rho_{\omega_1\dots\omega_N} |\Omega_-(s)\rangle \langle \Omega_0(s)|
-
\frac 1 4
[|\Omega_-(s)\rangle\langle \Omega_-(s)|,\rho_{\omega_1\dots\omega_N}]_+
\right)
\nonumber
\\
&+&
\gamma(\Omega_+(s) - \Omega_-(s))
\frac{\beta^2}{2}
\nonumber\\
&\pp+&\times
\left(
\frac 1 2
|\Omega_-(s)\rangle \langle \Omega_+(s)| \rho_{\omega_1\dots\omega_N} |\Omega_+(s)\rangle \langle \Omega_-(s)|
-
\frac 1 4
[|\Omega_+(s)\rangle\langle \Omega_+(s)|,\rho_{\omega_1\dots\omega_N}]_+
\right),
\label{eq:me2}
\end{eqnarray}
and, up to the presence of $s/N$ in terms involving $\alpha^2$, is identical to the last equation from the Appendix of \cite{pra}. In consequence, we can directly apply the results from \cite{pra} to (\ref{eq:me2}). One of the consequences of (\ref{eq:me2}) is the time-independence of
\be
p_{\omega_1\dots\omega_N}
&=&
\tr \rho_{\omega_1\dots\omega_N}(t)
=
\tr \pi_{\omega_1\dots\omega_N}\rho(t)\nonumber\\
&=&
\tr \Pi_{\omega_1\dots\omega_N}\rho(t)
\ee
$p_{\omega_1\dots\omega_N}$ is the probability of finding the sequence ${\omega_1\dots\omega_N}$ if one randomly and independently selects each $\omega$. The probability of finding $\omega$ equals $Z_\omega=|O_\omega|^2$, so $p_{\omega_1\dots\omega_N}=Z_{\omega_1}\dots Z_{\omega_N}$. Now, let
\be
\varrho_{\omega_1\dots\omega_N}(t)=\rho_{\omega_1\dots\omega_N}(t)/p_{\omega_1\dots\omega_N}
\ee
be a normalized solution of (\ref{eq:me2}), with the initial condition
\begin{eqnarray}
\varrho_{\omega_1\dots\omega_N}(0)&=&
|e,0_{\omega_1\dots\omega_N}\rangle\langle e,0_{\omega_1\dots\omega_N}|
\nonumber\\
&=&
\frac 1 2
\Big(
|\Omega_+(s)\rangle \langle \Omega_+(s)|
+
|\Omega_-(s)\rangle \langle \Omega_-(s)|
-
|\Omega_+(s)\rangle \langle \Omega_-(s)|
-
|\Omega_-(s)\rangle \langle \Omega_+(s)|
\Big).\nonumber\\
\end{eqnarray}
The parameters $\gamma_1=\gamma(\Omega_+(s) - \Omega_0(s)) \alpha^2$,
$\gamma_2=\gamma(\Omega_-(s) - \Omega_0(s))\alpha^2$, and
$\gamma_3=\gamma(\Omega_+(s) - \Omega_-(s)) \beta^2/2$ are related to the system-reservoir interaction Hamiltonian in a way that is identical to what was found in \cite{pra}. In order to have a well defined limit $N\to\infty$ we assume they are independent of $s$. However, in (\ref{eq:me2})  $\gamma_1$ and $\gamma_2$ are additionally multiplied by $s/N$, a fact that introduces an $s$-dependence into $\gamma_1(s)=\gamma_1 s/N$, $\gamma_2(s)=\gamma_2 s/N$, keeping $\gamma_3$ independent of $s$. Applying the damping-basis method (see the Appendix), we find
\begin{eqnarray}
\varrho_{\omega_1\dots\omega_N}(t)
&=&
-
\frac 1 2
\frac{ 1  }{ \gamma_1(s) - \gamma_2(s) + \gamma_3}
e^{- \frac{ \gamma_1(s) + \gamma_3 }{ 2 } t}
\nonumber
\\
&&
\times
\bigg(
- (\gamma_1(s) - \gamma_2(s) + \gamma_3)  |\Omega_+(s)\rangle \langle \Omega_+(s)|
+
\gamma_3   |\Omega_-(s)\rangle \langle \Omega_-(s)|\nonumber\\
&&
\pp{\bigg(+-}
+
( \gamma_1(s) - \gamma_2(s) )   |\Omega_0(s)\rangle \langle \Omega_0(s)|
\bigg)
\nonumber
\\
&+&
\frac 1 2
\frac{ \gamma_1(s) - \gamma_2(s) + 2 \gamma_3 }{\gamma_1(s) - \gamma_2(s) + \gamma_3 }
e^{- \frac{ \gamma_2(s) }{ 2 } t}
\Big(
|\Omega_-(s)\rangle \langle \Omega_-(s)|
-
|\Omega_0(s)\rangle \langle \Omega_0(s)|
\Big)
\nonumber
\\
&+&
|\Omega_0(s)\rangle \langle \Omega_0(s)|
\nonumber
\\
&-&
\frac 1 2
e^{- 2 i {q} \sqrt{s/N} t }
e^{- \frac{ \gamma_1(s) + \gamma_2(s) + \gamma_3 }{4} t}
|\Omega_+(s)\rangle \langle \Omega_-(s)|
\nonumber
\\
&-&
\frac 1 2
e^{2 i {q} \sqrt{s/N} t }
e^{- \frac{ \gamma_1(s) + \gamma_2(s) + \gamma_3 }{4} t}
|\Omega_-(s)\rangle \langle \Omega_+(s)|.
\end{eqnarray}
The conditional probability of finding the atom in its ground state, under the condition that the sequence is ${\omega_1\dots\omega_N}$, reads
\begin{eqnarray}
p_{g}(s,t)
&=&
\tr |g\rangle \langle g|\varrho_{\omega_1\dots\omega_N}(t)
\nonumber
\\
&=&
1
-
\frac 1 4
\frac{ \gamma_1(s) - \gamma_2(s) + 2\gamma_3 }{ \gamma_1(s) - \gamma_2(s) + \gamma_3 }
e^{- \frac{ \gamma_2(s) }{ 2 }  t}
-
\frac 1 4
\frac{ \gamma_1(s) - \gamma_2(s) }{ \gamma_1(s) - \gamma_2(s) + \gamma_3 }
e^{- \frac{ \gamma_1(s) + \gamma_3 }{ 2 } t}
\nonumber\\
&\pp=&-
\frac 1 2
e^{- \frac{ \gamma_1(s) + \gamma_2(s) + \gamma_3 }{4} t}
\cos 2  {q} \sqrt{s/N} t.
\label{eq:70}
\end{eqnarray}
It is instructive to confront the formula (\ref{eq:70}) with the one we would have obtained in the formalism from \cite{pra} if, at $T=0$, we employed an irreducible representation with $a_0={\cal{Z}}1$, for some constant ${\cal{Z}}>0$, i.e. with $[a,a^\dagger]={\cal{Z}}1$ (if ${\cal Z}<0$, then $a$ is a creation operator).
We first define $\tilde a=a/\sqrt{\cal Z}$, $\tilde a^\dag=a^\dag/\sqrt{\cal Z}$ and then perform calculations with $H$ expressed in terms of these new operators and appropriately rescaled parameters (see the Appendix). The probability is \cite{eps=0}
\begin{eqnarray}
p^{\text{irr}}_g(t)
&=&
1
-
\frac 1 4
\frac{ \gamma_1 {\cal{Z}} - \gamma_2{\cal{Z}} + 2\gamma_3 }{ \gamma_1{\cal{Z}} - \gamma_2{\cal{Z}} + \gamma_3 }
e^{- \frac{ \gamma_2{\cal{Z}} }{ 2 }  t}
-
\frac 1 4
\frac{ \gamma_1{\cal{Z}} - \gamma_2{\cal{Z}} }{ \gamma_1{\cal{Z}} - \gamma_2{\cal{Z}} + \gamma_3 }
e^{- \frac{ \gamma_1{\cal{Z}} + \gamma_3 }{ 2 } t}
\nonumber\\
&\pp=&-
\frac 1 2
e^{- \frac{ \gamma_1{\cal{Z}} + \gamma_2{\cal{Z}} + \gamma_3 }{4} t}
\cos 2  {q} \sqrt{{\cal{Z}}} t.
\label{eq:220}
\end{eqnarray}
Obviously, $\cal Z$ occurs in (\ref{eq:220}) in the same place as $s/N$ in (\ref{eq:70}). This is consistent with the fact that in subspaces characterized by $s$, the right-hand side of an analogous commutator involves $s/N$. All irreducible representations imply the same physical result provided one defines {\it observable parameters\/} by their renormalized forms: $q_{\text{ph}}={q} \sqrt{\cal{Z}}$, $\gamma_{1,\text{ph}}=\gamma_1\cal{Z}$, $\gamma_{2,\text{ph}}=\gamma_2\cal{Z}$, $\gamma_{3,\text{ph}}=\gamma_3$.

Returning to the reducible representation, the ground-state probability we are looking for is the weighted sum
\begin{eqnarray}
p_g(t)
&=&
\sum_{\omega_1\dots \omega_N}
p_{\omega_1\dots \omega_N}
\tr  |g\rangle \langle g|\varrho_{\omega_1\dots\omega_N}(t)\nonumber\\
&=&
\sum_{s=0}^N
{N \choose s}
Z_\omega^s
(1-Z_\omega)^{N-s}
p_g(s,t).
\label{eq:200}
\end{eqnarray}
This is basically the final formula that should be compared with experiment.
Before we do that, however, we have to relate the bare parameters $\gamma_1$, $\gamma_2$, $\gamma_3$, and $q$, to  their physical, renormalized counterparts.

In order to do so, we consider the asymptotic limit $N\to\infty$. The weak law of large numbers (Feller's theorem \cite{Feller,Feller2}) implies
\begin{eqnarray}
\lim_{N\to \infty}
p_g(t)
&=&
1
-
\frac 1 4
\frac{ \gamma_1 Z_\omega - \gamma_2Z_\omega + 2\gamma_3 }{ \gamma_1Z_\omega - \gamma_2Z_\omega + \gamma_3 }
e^{- \frac{ \gamma_2Z_\omega }{ 2 }  t}
-
\frac 1 4
\frac{ \gamma_1Z_\omega - \gamma_2Z_\omega }{ \gamma_1Z_\omega - \gamma_2Z_\omega + \gamma_3 }
e^{- \frac{ \gamma_1Z_\omega + \gamma_3 }{ 2 } t}
\nonumber\\
&\pp=&
-
\frac 1 2
e^{- \frac{ \gamma_1Z_\omega + \gamma_2Z_\omega + \gamma_3 }{4} t}
\cos 2  {q} \sqrt{Z_\omega} t.
\label{eq:210}
\end{eqnarray}
The result is practically identical to (\ref{eq:220}). ${q} \sqrt{Z_\omega}$ is the effective physical coupling, where $Z_\omega=|O_\omega|^2$ is the probability of finding the mode $\omega$. Defining $Z=\max_\omega\{Z_\omega\}$, we first of all note that the formula involves an automatic cut-off $\chi_\omega=Z_\omega/Z$, $0\leq \chi_\omega\leq 1$. The observable parameters are identified with
$q_{\text{ph}}={q} \sqrt{Z}$, $\gamma_{1,\text{ph}}=\gamma_1Z$, $\gamma_{2,\text{ph}}=\gamma_2Z$, and $\gamma_{3,\text{ph}}=\gamma_3$.
The reducible-representation asymptotic formula reconstructs {\it exactly\/} the standard one following from the analysis given in \cite{pra} if we assume that $\chi_\omega= 1$ for frequencies belonging to the optical regime.
The formula for $p_g(t)$, valid for {\it all\/} values of $1\leq N<\infty$, finally becomes
\begin{eqnarray}
p_g(t)
&=&
\sum_{s=0}^N
{N \choose s}
Z^s
(1-Z)^{N-s}
\times
\nonumber
\\
&&
\times
\left\{
1
-
\frac 1 4
\frac{ \frac{s}{N Z} \left( \gamma_{1,\text{ph}}  - \gamma_{2,\text{ph}} \right) + 2\gamma_3 }
{ \frac{s}{N Z} \left( \gamma_{1,\text{ph}}  - \gamma_{2,\text{ph}} \right) + \gamma_3 }
e^{- \frac{ \gamma_{2,\text{ph}} \frac{s}{N Z} }{ 2 }  t}
-
\frac 1 4
\frac{ \frac{s}{N Z} \left(  \gamma_{1,\text{ph}}  - \gamma_{2,\text{ph}} \right) }
{ \frac{s}{N Z} \left( \gamma_{1,\text{ph}}  - \gamma_{2,\text{ph}} \right) + \gamma_3 }
e^{- \frac{ \gamma_{1,\text{ph}} \frac{s}{N Z} + \gamma_3 }{ 2 } t}
\right.
\nonumber
\\
&&
\left.
-
\frac 1 2
e^{- \frac{  \frac{s}{N Z} \left( \gamma_{1,\text{ph}} + \gamma_{2,\text{ph}} \right) + \gamma_3 }{4} t}
\cos 2   {q}_{\text{ph}}  \sqrt{\frac{ s }{ N Z }}  t
\right\}.
\nonumber
\\
\label{eq:pgt}
\end{eqnarray}
The limit $\lim_{N\to\infty}p_g(t)$ reconstructs (\ref{eq:220}) with ${\cal Z}=Z$. The larger $N$, the less important the exact value of $Z$. For $N$ of the order of $10^5$ or higher, plots of $p_g(t)$ are insensitive to changes of $Z$ if the product $\varsigma=NZ$ is kept constant. Such a limit, $N\to\infty$, $\varsigma=$const, is precisely a thermodynamic limit with fixed effective number $\varsigma$ of oscillators that interact with the two-level system [compare Eq.~(\ref{int of stigma})]. The thermodynamic limit implicitly removes the cut-off since $Z=\varsigma/N\to 0$, $Z=\max_\omega\{Z_\omega\}$, $\sum_\omega Z_\omega=1$, implies shifting cut-off to infinity.   Requirements of mathematical consistency imply that $N$ is finite. For physical reasons, however, $N$ must be very large, and thus $Z$ is small but non-zero.

Let us now assume that the cavity is identical to the one employed in \cite{Brune}. The mode has a Gaussian structure, so we have to correct (\ref{eq:pgt}) in a way described in detail in \cite{pra}. Denoting by $d$ and $w$ the cavity length and the Gaussian width, respectively, we get the effective probability
\begin{eqnarray}
\tilde p_g(t)
&=&
\sum_{s=0}^N
{N \choose s}
Z^s
(1-Z)^{N-s}
\times
\nonumber
\\
&&
\times
\left\{
1
-
\frac 1 4
\frac{ \frac{s}{N Z} \left( \gamma_{1,\text{ph}}  - \gamma_{2,\text{ph}} \right) + 2\gamma_3 }
{ \frac{s}{N Z} \left( \gamma_{1,\text{ph}}  - \gamma_{2,\text{ph}} \right) + \gamma_3 }
e^{- \frac{ \gamma_{2,\text{ph}} \frac{s}{N Z} }{ 2 \sqrt{\pi}w/d}  t}
-
\frac 1 4
\frac{ \frac{s}{N Z} \left(  \gamma_{1,\text{ph}}  - \gamma_{2,\text{ph}} \right) }
{ \frac{s}{N Z} \left( \gamma_{1,\text{ph}}  - \gamma_{2,\text{ph}} \right) + \gamma_3 }
e^{- \frac{ \gamma_{1,\text{ph}} \frac{s}{N Z} + \gamma_3 }{ 2\sqrt{\pi}w/d } t}
\right.
\nonumber
\\
&&
\left.
-
\frac 1 2
e^{- \frac{  \frac{s}{N Z} \left( \gamma_{1,\text{ph}} + \gamma_{2,\text{ph}} \right) + \gamma_3 }{4\sqrt{\pi}w/d} t}
\cos 2   {q}_{\text{ph}}  \sqrt{\frac{ s }{ N Z }}  t
\right\},
\nonumber
\\
\label{eq:pgt-eff}
\end{eqnarray}
where $t$ in (\ref{eq:pgt-eff}) is the {\it effective time\/} \cite{pra}. Analogously, the Gaussian-mode correction to the irreducible case reads
\begin{eqnarray}
\tilde p^{\text{irr}}_g(t)
&=&
1
-
\frac 1 4
\frac{ \gamma_1 {\cal{Z}} - \gamma_2{\cal{Z}} + 2\gamma_3 }{ \gamma_1{\cal{Z}} - \gamma_2{\cal{Z}} + \gamma_3 }
e^{- \frac{ \gamma_2{\cal{Z}} }{ 2 \sqrt{\pi}w/d}  t}
-
\frac 1 4
\frac{ \gamma_1{\cal{Z}} - \gamma_2{\cal{Z}} }{ \gamma_1{\cal{Z}} - \gamma_2{\cal{Z}} + \gamma_3 }
e^{- \frac{ \gamma_1{\cal{Z}} + \gamma_3 }{ 2 \sqrt{\pi}w/d} t}
\nonumber\\
&\pp=&-
\frac 1 2
e^{- \frac{ \gamma_1{\cal{Z}} + \gamma_2{\cal{Z}} + \gamma_3 }{4\sqrt{\pi}w/d} t}
\cos 2  {q} \sqrt{{\cal{Z}}} t.
\label{eq:220-eff}
\end{eqnarray}
Fig.~1 shows the time-dependence of $\tilde p_g(t)$ for $N=10^5$  and various values of $\varsigma$. Even more suggestive is the plot of the difference $|\tilde p_g^{\rm irr}(t)-\tilde p_g(t)|$ (Fig.~2) compared with the error bars taken from the data of Brune {\it et al.\/} \cite{Brune}. It is clear that this concrete experiment cannot discriminate between the limit $N\to\infty$ (i.e. the standard theory based on irreducible representations) and the alternative non-Wightmanian theory with any finite $\varsigma>400$.
\begin{center}
\begin{figure}[t]
\includegraphics[width=1.0\textwidth]{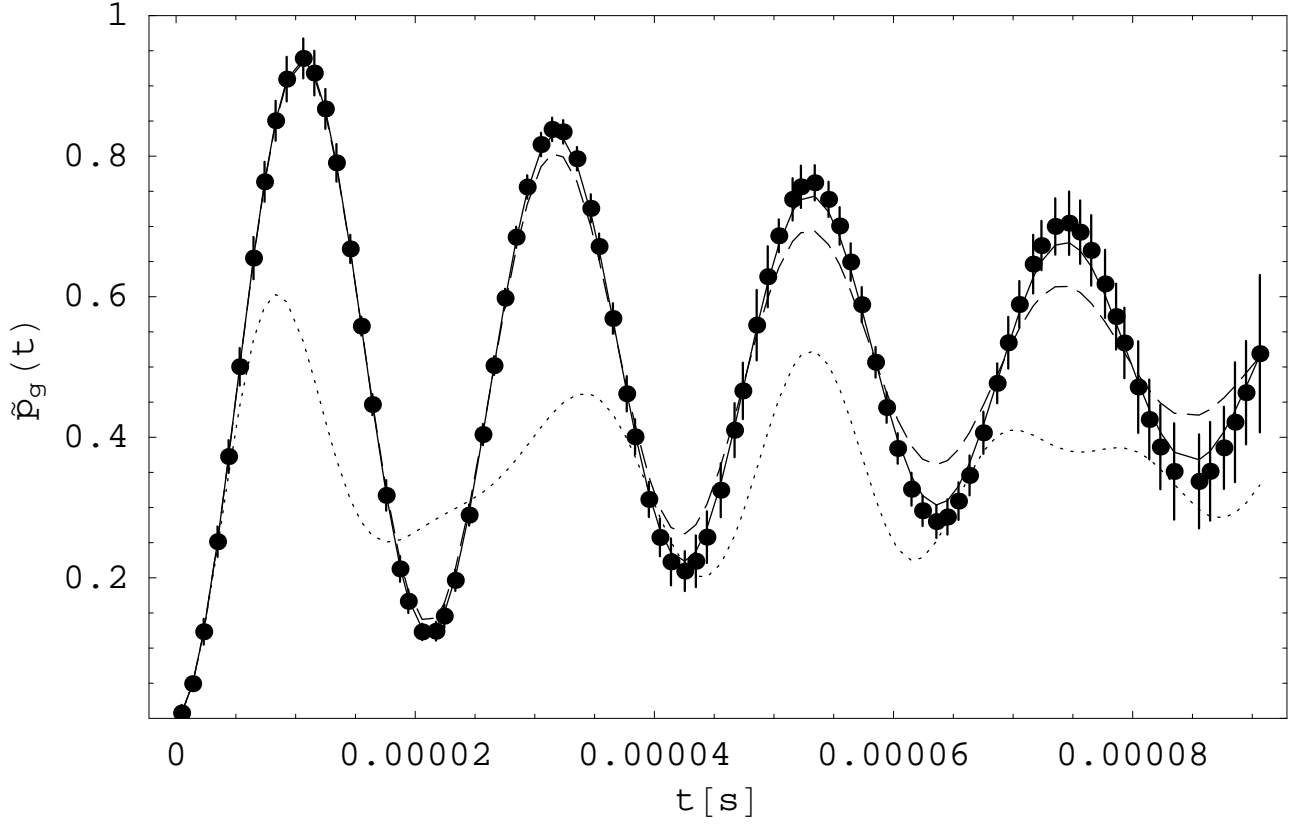}%
\caption{Probability $\tilde p_g(t)$, Eq.~(\ref{eq:pgt-eff}), of finding the atom in the lower state $|g\rangle$ for $\varsigma=1$ (dotted), $\varsigma=100$ (dashed), and $\varsigma=400$ (solid), with $N=10^5$ for all the three cases. $t$ is the effective time. The filled circles represent an analogous probability obtained for irreducible representations. Error bars taken from the experiment of Brune {\it et al.\/} provide a natural measure of distance between predictions of the alternative theories. For higher values of $\varsigma$, say $\varsigma=1000$, the reducible representation becomes indistinguishable from the irreducible one. The remaining parameters are: ${q}=47 \pi 10^3$ Hz, $\gamma_1=\gamma_2=83.912$ Hz and $\gamma_3=0.07 {q}$.}
\label{fig:pg-various-NZ}
\end{figure}

\begin{figure}[t]
\includegraphics{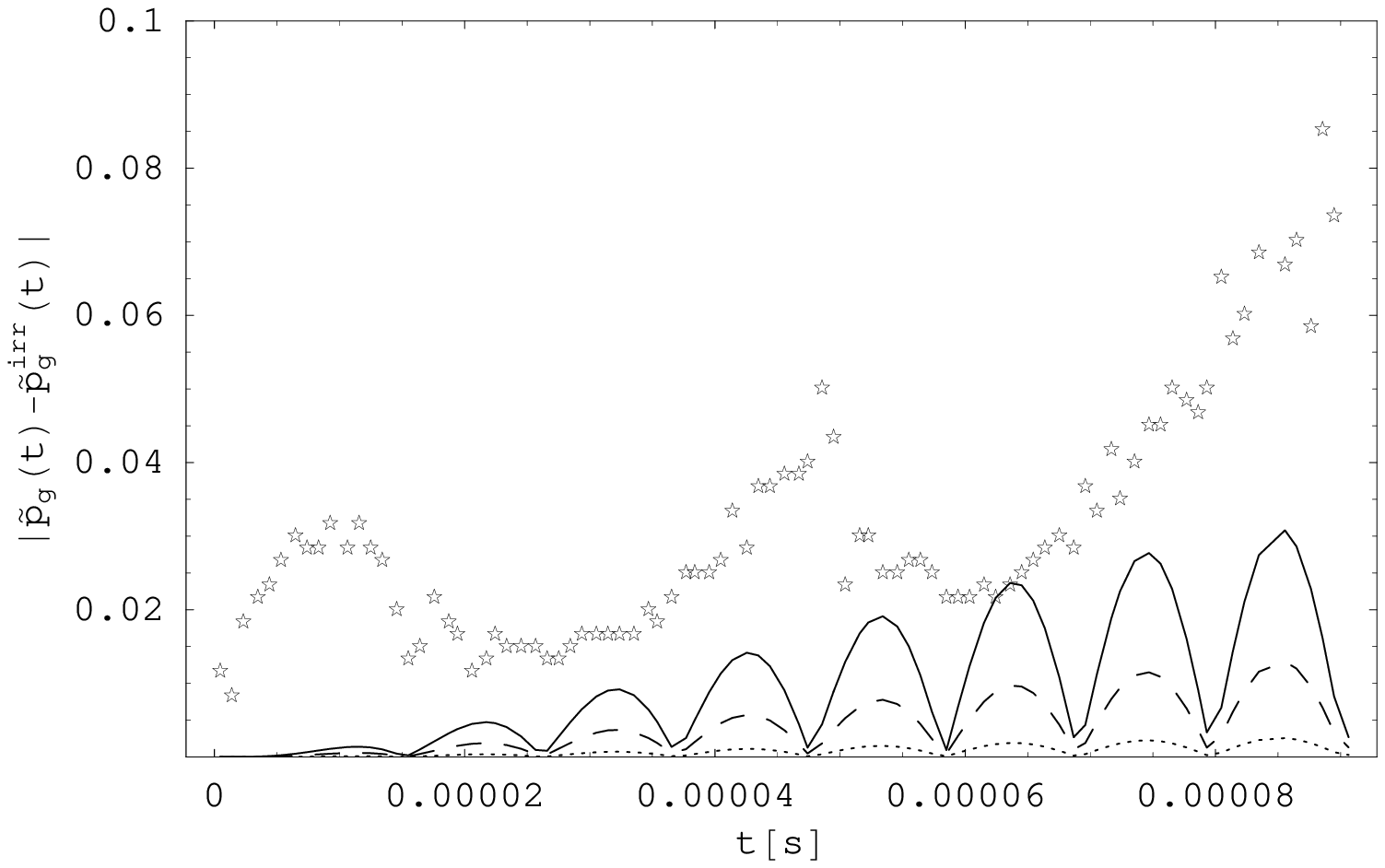}%
\caption{The difference $|\tilde p_g^{\rm irr}(t)-\tilde p_g(t)|$, with $\tilde p_g^{\rm irr}(t)$ and $\tilde p_g(t)$ given by (\ref{eq:220-eff}) and (\ref{eq:pgt-eff}), respectively.  $t$ is the effective time.
$\varsigma=400$ (solid), $\varsigma=1000$ (dashed), $\varsigma=5000$ (dotted), $N=10^5$. Stars represent the error bars taken from the experiment of Brune {\it et al.\/}  The curves remain  practically unchanged for higher $N$, so the plots survive the thermodynamic limit with $\varsigma=$~const.}
\label{fig:pg-various-NZ1}
\end{figure}

\end{center}

\section{Energy decay}

Identification of physical parameters with the renormalized ones is supported by the analysis of energy losses. In irreducible representations with $[a,a^\dagger]={\cal{Z}}1$, the average energy  of the atom-field system inside of the cavity, at $T=0$, $E(t)=\hbar \overline{\Omega(t)}{^{\text{irr}}}=\hbar \tr \Omega\rho(t)$ [see the Appendix for the explicit form of $\rho(t)$], is given by
\begin{eqnarray}
\overline{\Omega(t)}{^{\text{irr}}}
&=&
-
\frac \omega 2
\nonumber
\\
&\pp=&
+
\frac 1 2
\left\{
\omega
\frac{ \gamma_1{\cal{Z}} - \gamma_2{\cal{Z}}  }{ \gamma_1{\cal{Z}} - \gamma_2{\cal{Z}} + \gamma_3}
+
{q\sqrt{{\cal{Z}}}}
\frac{ \gamma_1{\cal{Z}} - \gamma_2{\cal{Z}} + 2 \gamma_3   }{ \gamma_1{\cal{Z}} - \gamma_2{\cal{Z}} + \gamma_3}
\right\}
e^{- \frac{ \gamma_1{\cal{Z}} + \gamma_3 }{ 2 } t}
\nonumber
\\
&\pp=&
+
\frac 1 2
\frac{ \gamma_1{\cal{Z}} - \gamma_2{\cal{Z}} + 2 \gamma_3 }{\gamma_1{\cal{Z}} - \gamma_2{\cal{Z}} + \gamma_3 }
\left(
\omega - {q\sqrt{{\cal{Z}}}}
\right)
e^{- \frac{ \gamma_2{\cal{Z}} }{ 2 } t}.\label{E1}
\end{eqnarray}
In \cite{pra} we showed that best fits to experimental data are found for $\gamma_1=\gamma_2$. Inserting
$\gamma=\gamma_1=\gamma_2$ into (\ref{E1}), we get
\be
\overline{\Omega(t)}{^{\text{irr}}}
&=&
-
\frac \omega 2
+\omega
e^{- \frac{ \gamma{\cal{Z}} }{ 2 } t}
+
q\sqrt{\cal Z}e^{- \frac{ \gamma{\cal{Z}} }{ 2 } t}\big(e^{- \frac{ \gamma_3 }{ 2 } t}-1\big).
\ee
It is evident that the energy damping parameter is $\gamma{\cal Z}$, and not just $\gamma$.

The reducible-representation result is similar,
\begin{eqnarray}
\tr \Omega \varrho_{\omega_1\ldots\omega_N}(t)
&=&
-
\frac \omega 2
\nonumber
\\
&&
+
\frac 1 2
\left\{
\omega
\frac{ \gamma_1(s) - \gamma_2(s)  }{ \gamma_1(s) - \gamma_2(s) + \gamma_3}
+
{q} \sqrt{ \frac s  N }
\frac{ \gamma_1(s) - \gamma_2(s) + 2 \gamma_3   }{ \gamma_1(s) - \gamma_2(s) + \gamma_3}
\right\}
e^{- \frac{ \gamma_1(s) + \gamma_3 }{ 2 } t}
\nonumber
\\
&&
+
\frac 1 2
\frac{ \gamma_1(s) - \gamma_2(s) + 2 \gamma_3 }{\gamma_1(s) - \gamma_2(s) + \gamma_3 }
\left(
\omega - {q} \sqrt{\frac s N }
\right)
e^{- \frac{ \gamma_2(s) }{ 2 } t}\\
&=&\overline{\Omega(s,t)}.
\end{eqnarray}
As before, the right-hand side depends on $s$ and not on the exact form of the sequence $\omega_1\ldots\omega_N$. Repeating the reasoning from the previous sections, we find
\be
\overline{\Omega(t)}
&=&
\sum_{s=0}^N
{N \choose s}
Z^s
(1-Z)^{N-s}
\overline{\Omega(s,t)}.
\ee
Fig.~3 compares the two expressions for various values of $\varsigma$, after having renormalized the parameters.
\begin{figure}[t]
\includegraphics[width=1.0\textwidth]{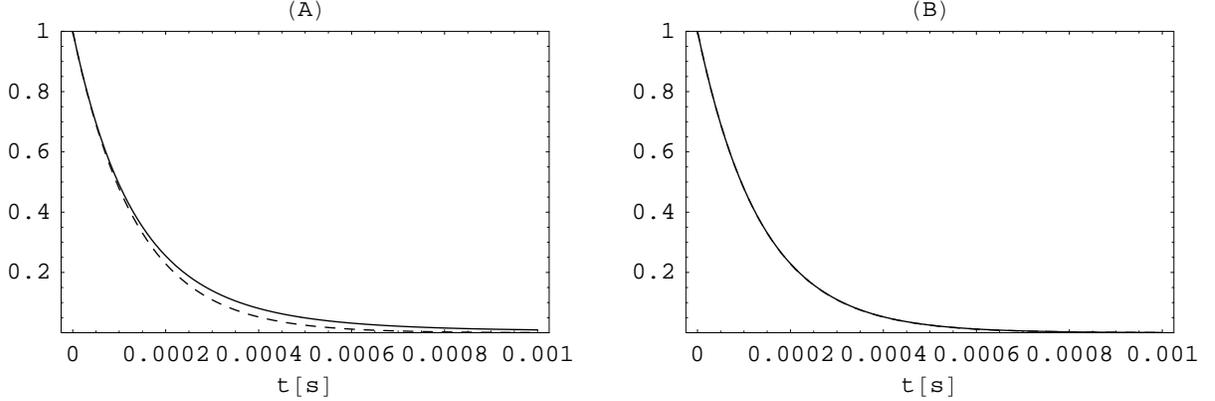}%
\caption{Comparison of $\overline{\Omega(t)}{^{\rm irr}}$ (dashed) and $\overline{\Omega(t)}$ (solid), for (A) $\varsigma=10$ and (B) $\varsigma=400$ (curves are indistinguishable). The other parameters are: $N=10^5$, $\gamma_{1,\rm ph}=\gamma_{2,\rm ph}=0.1 {q_{\rm ph}}$, $\gamma_3=0.001 {q_{\rm ph}}$, ${q_{\rm ph}}=47 \pi 10^3$ Hz.}
\label{fig:average-energy}
\end{figure}

\section{Vacuum collapses and revivals for $\varsigma<\infty$}

The discussed Rabi-oscillation data do not distinguish between the two alternative forms of field quantization: one can always increase the value of $\varsigma$ and produce a theory indistinguishable from the standard one within some given error bars. However, monitoring the oscillation long enough we can determine the value of $\varsigma$, provided $\varsigma<\infty$.

\begin{figure}[t]
\includegraphics[width=0.6\textwidth]{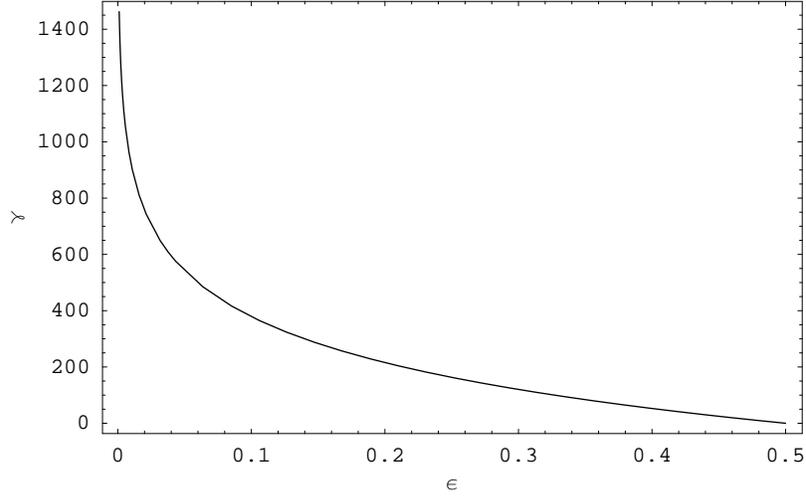}%
\caption{Plot of $\gamma=\gamma_{1,\text{ph}} + \gamma_{2,\text{ph}} + \gamma_3$ as a function of $\varepsilon$. In order to observe the revival, the value of $\varepsilon$ should be greater than experimental errors.}
\label{fig:gamma-epsilon}
\end{figure}

\begin{figure}[t]
\includegraphics[width=1.0\textwidth]{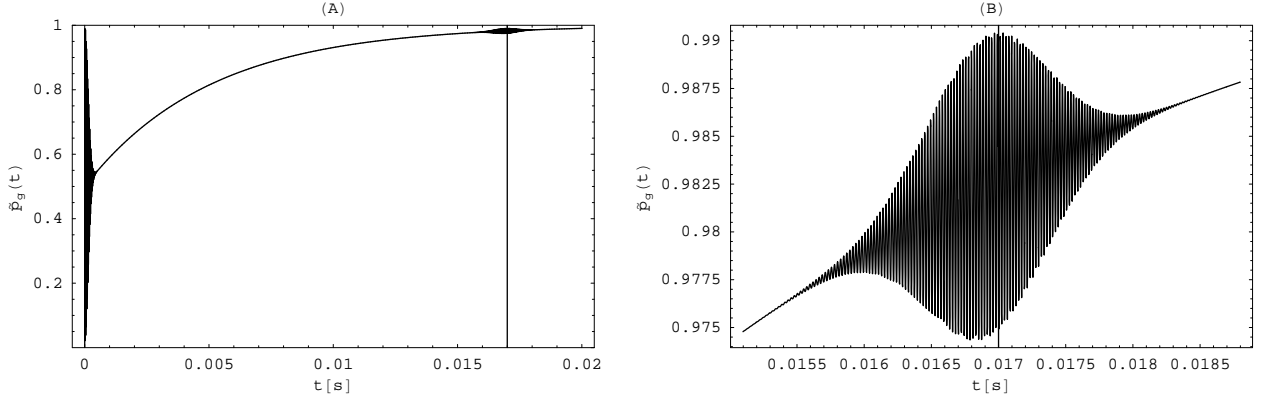}%
\caption{Probability of the atomic ground state as a function of the effective time $t$ for $\gamma_1=\gamma_2=83.912$ Hz, $\gamma_3=10$ Hz, and $\varsigma=400$ (these parameters yield $\varepsilon\approx 0.23$). The revival time (\ref{t_r}), $t_r\approx 0.017$~s, is indicated by the vertical line. It approximately determines the moment of maximal visibility of the revival.}
\label{fig:revival}
\end{figure}

For a finite $N$, the eigenvalues $s/N$ are distributed around the most probable value $s/N\approx Z$ (i.e. $s\approx \varsigma$), resulting in distribution of Rabi frequencies even in exact vacuum. In consequence, instead of a single-frequency oscillation we obtain beats analogous to those occurring in Rabi oscillations in the presence of a coherent light. Rabi frequency in exact resonance and in a subspace characterized by $s$ equals
\be
2q\sqrt{\frac{s}{N}}
&=&
2q\sqrt{Z}\sqrt{\frac{s}{NZ}}
=
2q_{\rm ph}\sqrt{\frac{s}{\varsigma}},
\ee
so that the most probable Rabi frequency is $2q_{\rm ph}$.
The first revival of a collapsed vacuum Rabi oscillation will have its maximum
when phases of neighboring and dominating oscillating terms differ by the factor of $2\pi$.
This means that the revival time, $t_r$, can be determined from
\begin{eqnarray}
2  {q}_{\text{ph}}   t_r
-
2 {q}_{\text{ph}}  \sqrt{ \frac{\varsigma-1}{\varsigma}  } t_r
&=&
2 \pi.
\end{eqnarray}
We find
\begin{eqnarray}
t_r
&=&
\big( \varsigma+ \sqrt{ \varsigma(\varsigma-1)}\big)\pi/{q}_{\text{ph}}.\label{t_r}
\end{eqnarray}
If we take dissipation into account, the revival can be seen only if the oscillation occurring in (\ref{eq:pgt}) is still visible.
The amplitude of oscillation is described, at $s\approx\varsigma$, by
\begin{eqnarray}
\varepsilon
=
\frac 1 2
e^{ - \frac{ \gamma_{1,\text{ph}} + \gamma_{2,\text{ph}} + \gamma_3 }{ 4 } t_r},
\end{eqnarray}
or
\begin{eqnarray}
\gamma &=&\gamma_{1,\text{ph}} + \gamma_{2,\text{ph}} + \gamma_3 \nonumber\\
&=&
-
\frac 4 t_r
\ln 2 \varepsilon
\end{eqnarray}
with $0<\varepsilon<1/2$. Actually, due to interference effects the amplitude of the revival is some ten times smaller than $\varepsilon$. The dependence of $\gamma$ on $\varepsilon$ is shown in Fig.~\ref{fig:gamma-epsilon}. Fig.~\ref{fig:revival} shows the revival of the decayed $p_g(t)$. The parameters used in Fig.~\ref{fig:pg-various-NZ} imply $\varepsilon \sim 10^{-20}$, so the effect would not be visible in experiments where error bars grow with time similarly to those from \cite{Brune}. Moreover, the data monitor oscillations for four Rabi periods, whereas for $\varsigma=400$, assuming the most optimistic scenario, the first revival should be seen after approximately $800\,T_{\rm Rabi}$.
\section{Finite $N$ or $N\to\infty$?}

Cavity boundary conditions imply that there are infinitely many $\omega$s. The probability of finding a given $\omega$ is
$Z_\omega=|O_\omega|^2$. Physical intuition suggests that $Z=\max_\omega\{|O_\omega|^2\}$ should be a small but non-zero number. Summability of probabilities to 1 implies that $Z_\omega\to 0$ if $\omega\to\infty$. On the other hand, evidently $Z_\omega\approx Z$ for optical frequencies since $\chi_\omega=Z_\omega/Z$ plays a role of a cut-off which, in the optical regime, should satisfy $\chi_\omega\approx 1$ (see, for example, a discussion of this point in \cite{MPMC}).

One of the important properties of field quantization in terms of reducible representations is the correspondence principle with standard regularized quantum optics  (it is beyond the scope of the present paper, but computation of, say, resonance fluorescence in the reducible-representation formalism indeed introduces $\chi_\omega$ in those places where one puts cut-off by hand in the standard formalism). What is interesting, the role of correspondence principle is played by the weak law of large numbers.

The weak law, $N\to\infty$ with $Z_\omega=$~const, replaces all $s/N$ by probabilities $Z_\omega$. At the level of representation, the weak law follows from spectral representations of central elements $a_0(\omega)=\uu I_\omega$ occurring at right-hand sides of the commutator $[\uu a_\omega,\uu a_{\omega'}^\dag]=\delta_{\omega\omega'}\uu I_\omega$ (recall that $\uu I_\omega$ are the frequency-of-success operators known from quantum laws of large numbers). The reducible representation may be regarded as a ``quantized" form of the standard {\it naively\/} regularized irreducible representation $[a_\omega,a_{\omega'}^\dag]=\delta_{\omega\omega'}Z_\omega 1$, where $Z_\omega$ is a regularizing function. The naive regularization is known to be in conflict with Poincar\'e covariance of canonical commutation relations. The replacement of the function $Z_\omega$ by the operator $\uu I_\omega$ leads to the correct behavior of the commutator under Poincar\'e transformations \cite{MCKW,MCJN,II,III}.

The thermodynamic limit, $N\to\infty$ with $\varsigma=NZ=$~const is different. Since $Z=\varsigma/N$, the bare charge satisfies
\be
q=\frac{q_{\rm ph}}{\sqrt{Z}}=\sqrt{\frac{N}{\varsigma}}q_{\rm ph}.
\ee
The Jaynes-Cummings interaction [see (\ref{uu CCR1})]
\be
{q}\sigma_+\uu a_\omega+{\rm H.c.}
&=&
\sqrt{\frac{N}{\varsigma}}q_{\rm ph}\sigma_+
\frac{1}{\sqrt{N}}
\Big(
a_\omega \otimes {I} \otimes \ldots \otimes {I}
+
\ldots
+
{I} \otimes \ldots \otimes {I} \otimes a_\omega
\Big)
+{\rm H.c.}\nonumber\\
&=&
\frac{q_{\rm ph}}{\sqrt{\varsigma}}
\sigma_+
\Big(
a_\omega \otimes {I} \otimes \ldots \otimes {I}
+
\ldots
+
{I} \otimes \ldots \otimes {I} \otimes a_\omega
\Big)
+{\rm H.c.}\label{int of stigma}
\ee
shows that the physical coupling parameter is, in fact, independent of $N$:
\be
\frac{q_{\rm ph}}{\sqrt{\varsigma}}
=
\frac{q}{\sqrt{N}},
\ee
for any $N$. Let us stress, that although we do not yet have exact results on the existence of the thermodynamic limit, all numerical experiments show quick convergence, with growing $N$, of $p_g(t)$ to a function whose shape is characterized by $\varsigma$.

Thermodynamic limits typically make some quantities divergent whereas some other quantities are well defined (think of a glass of water --- the density of water will correctly behave in the limit, but the water mass will become infinite). A similar situation is encountered in our treatment of cavity QED. The probability $p_g(t)$ becomes well defined even if $N$ tends to infinity with $\varsigma$ kept constant, but the vacuum energy diverges.
So, physically we have to assume that $N$ is large but finite, similarly to $Z$ which is small but nonzero.

\section{Overview and conclusions}

All quantum optical experiments suggest that the idea of replacing classically oscillating amplitudes, $\alpha_\omega e^{-i\omega t}$, by quantum harmonic-oscillator operators, $a_\omega e^{-i\omega t}$, is correct. What is important, the representations of harmonic-oscillator Lie algebras we know from quantum mechanics textbooks correspond physically to quantum oscillators whose $\omega$ is a classical {\it parameter\/}. (Heisenberg, Born, and Jordan  were not aware, in 1925, of the role of eigenvalues for quantum theories --- the point was understood by Schr\"odinger in 1926 \cite{Jammer}.) Simultaneously, for physical reasons the frequency $\omega$ is in actual oscillators a function of observables (for example, endpoints of a nano-scale pendulum are given by atomic center-of-mass wavepackets and, hence, its length is unsharp --- the pendulum exists in a quantum superposition of different lengths $l$ or, equivalently, of the corresponding frequencies $\omega=\sqrt{g/l}$).

Attempts of taking this observation into account lead to oscillator wave packets existing in superpositions of different $\omega$s. The frequencies are given by eigenvalues of an operator $\hat \omega$ that, in simplest models, commutes with other oscillator observables (in yet more realistic cases the operators $\hat\omega$ will not commute with other observables). Since all $\omega$s can be associated with a single oscillator, there is no relation between the number (typically infinite) of field modes and the number $N$ of oscillators (that can be finite). 

For those who are trained on the usual Heisenberg 1925 oscillator, the conclusion seems counterintuitive: Now a single oscillator may contain many modes, and a single-mode problem may involve many oscillators. An algebraic implication is that once we replace $\omega$ by $\hat\omega$, a new representation of the harmonic oscillator Lie algebra arises. The representation is reducible, and the commutator $[a_\omega,a_\omega^\dag]$ is no longer proportional to the identity, but is a nontrivial operator that commutes with all the operators of the algebra. The vacuum average of this operator plays a role of a regularizing function that, at least in all the applications considered so far, appears in exactly those places where one expects regularization to occur. Simultaneously, at the level of spectra, instead of regularizing functions $Z_\omega$ one finds the eigenvalues $s/N$, $s=0,1\dots N$, of the operator $[a_\omega,a_\omega^\dag]$. The eigenvalues $s/N$ are not distributed uniformly. The weak law of large numbers implies that for a large $N$ they concentrate around the probability $Z_\omega=|O_\omega|^2$ of finding the eigenvalue $\omega$ in vacuum. The maximal value $Z=\max_\omega \{Z_\omega\}$ is a parameter that characterizes the structure of the vacuum state. The product $\varsigma=ZN$ represents an effective number of field oscillators that interact with the two-level atom. Since for physical reasons $Z$ should be small, then keeping $\varsigma$ fixed we have to assume that $N$ is large. Actually, so large that the thermodynamic limit, with $\varsigma=\,$const, should be applicable. In relativistic version of the theory both $Z$ and $N$ are relativistically invariant. 

As we can see, there are reasons to regard the reducible representation of harmonic oscillator algebras as physically justified. Now we face two questions. First, can the available data really prove that the representation of the Lie algebra is the one we know from standard quantum optics? The answer is: Probably no, since the limit $N\to\infty$ is a correspondence principle with the standard theory, so one can always increase $N$ to approach the standard theory within given experimental error bars. So the second question is: Does there exist a phenomenon whose observation could prove that $N$ and $\varsigma$ are finite? The answer is, in principle, positive. For example, monitoring a decayed Rabi oscillation in exact vacuum we expect its revival after, approximately,  $2\varsigma T_{\rm Rabi}$. Current data suggest $\varsigma>400$. 

It is also worthy of mentioning that the parameters $Z$ and $N$ alone may not be measurable, since $Z$ gets hidden in the renormalized charge $e_{\rm ph}=e\sqrt{Z}$ (note the analogy between $Z$ and the renormalization constant $Z_3$). Once incorporated into renormalized parameters, $Z$ disappears from large-$N$ asymptotic formulas, but what remains are the ``cutoff" functions $\chi_\omega=Z_\omega/Z$. The thermodynamic limit turns out to remove the cutoff. In effect, the standard steps of the renormalization procedure are here interpretable in physical terms.
If we could confirm that physical $\varsigma$ and $N$ are finite, we would have a proof that physical vacua are more similar to realistic Bose-Einstein condensates than to ``unique and Poincar\'e invariant cyclic vectors" typical of Wightmanian axiomatics.

\acknowledgments
The work was supported by VUB, Brussels, and Polish national scientific network LFPPI.

\section*{Appendix A: $T=0$ solution in irreducible representations}

For $T=0.8$ K the proportionality factor $\epsilon$ linking transitions up and down between dressed states from the same manifold equals $\exp(-2\hbar q/kT)\approx 0.999\,997$, but vanishes for $T=0$ (compare Fig.~10 in \cite{pra}). The case $T=0$ was not interesting from the point of view of the analysis given in \cite{pra}, so we made there the approximation $\epsilon=1$. In consequence, an appropriate $T=0$ formula for $p_g^{\rm irr}(t)=\tr |g\rangle\langle g|\rho(t)$ is missing in \cite{pra}. Below we give a detailed derivation of the solution corresponding to $T=0$. This supplements \cite{pra}, simultaneously showing how to solve the master equations occurring in the present paper, since mathematically they are identical.

The irreducible representation is $[a,a^\dag]=1$; if $[a,a^\dag]={\cal Z}1$, ${\cal Z}>0$, we redefine $a/\sqrt{\cal Z}\to a$, $a^\dag/\sqrt{\cal Z}\to a^\dag$.
The $T=0$ master equation  reads
\begin{eqnarray}
\dot{\rho}
&=&
-i[\Omega,\rho]
+
\gamma_1
\left\{
\frac 1 2
|\Omega_0\rangle \langle \Omega_+| \rho |\Omega_+\rangle \langle \Omega_0|
-
\frac 1 4
\Big[| \Omega_+\rangle \langle \Omega_+|,\rho\Big]_+
\right\}
\nonumber
\\
&\pp=&
+
\gamma_2
\left\{
\frac 1 2
|\Omega_0\rangle \langle \Omega_-| \rho |\Omega_-\rangle \langle \Omega_0|
-
\frac  1 4
\Big[| \Omega_-\rangle \langle \Omega_-|,\rho\Big]_+
\right\}
\nonumber
\\
&\pp=&
+
\gamma_3
\left\{
\frac 1 2
|\Omega_-\rangle \langle \Omega_+| \rho |\Omega_+\rangle \langle \Omega_-|
-
\frac 1 4
\Big[| \Omega_+\rangle \langle \Omega_+|,\rho\Big]_+
\right\}.
\label{eq:meT0}
\end{eqnarray}
Eq.~(\ref{eq:meT0}), written as $\dot\rho={\cal L}\rho$, defines the operator $\cal L$. The eigenvectors ${\cal L}\rho_j=\Lambda_j\rho_j$ define the so-called damping bases \cite{Englert}:
\begin{subequations}
\begin{eqnarray}
\rho_1
&=&
- (\gamma_1 - \gamma_2 + \gamma_3)  |\Omega_+\rangle \langle \Omega_+|
+
\gamma_3   |\Omega_-\rangle \langle \Omega_-|
+
( \gamma_1 - \gamma_2 )   |\Omega_0\rangle \langle \Omega_0|,
\quad
\Lambda_1 \ = \ - \frac{  \gamma_1 + \gamma_3 }{ 2 },
\\
\rho_2
&=&
|\Omega_-\rangle \langle \Omega_-|
-
|\Omega_0\rangle \langle \Omega_0|,
\quad
\Lambda_2 \ = \ - \frac{ \gamma_2 }{ 2 },
\\
\rho_3
&=&
( \gamma_1 \gamma_2 + \gamma_2 \gamma_3 )
|\Omega_0\rangle \langle \Omega_0|,
\quad
\Lambda_3 \ = \ 0,\\
\rho_4
&=&
|\Omega_+\rangle \langle \Omega_-|,
\quad
\Lambda_4
\ = \
-i (\Omega_+ - \Omega_-)
-
\frac{ \gamma_1 + \gamma_2 + \gamma_3  }{4},
\\
\rho_5
&=&
|\Omega_+\rangle \langle \Omega_0|,
\quad
\Lambda_5
\ = \
-i (\Omega_+ - \Omega_0)
-
\frac{ \gamma_1 + \gamma_3 }{ 4 },
\\
\rho_6
&=&
|\Omega_-\rangle \langle \Omega_0|,
\quad
\Lambda_6
\ = \
-i(\Omega_- - \Omega_0 )
-
\frac{ \gamma_2 }{ 4 } ,
\\
\rho_7
&=&
|\Omega_-\rangle \langle \Omega_+|,
\quad
\Lambda_7
\ = \
i (\Omega_+ - \Omega_-)
-
\frac{ \gamma_1 + \gamma_2 + \gamma_3  }{4},
\\
\rho_8
&=&
|\Omega_0\rangle \langle \Omega_+|,
\quad
\Lambda_8
\ = \
i (\Omega_+ - \Omega_0)
-
\frac{ \gamma_1 + \gamma_3 }{ 4 },
\\
\rho_9
&=&
|\Omega_0\rangle \langle \Omega_-|,
\quad
\Lambda_9
\ = \
i(\Omega_- - \Omega_0 )
-
\frac{ \gamma_2 }{ 4 }.
\end{eqnarray}
\label{eq:basis1}
\end{subequations}
The initial condition
\begin{eqnarray}
\rho(0)
&=&
|e,0\rangle \langle e,0|
=
\frac 1 2 |\Omega_+\rangle \langle \Omega_+|
+
\frac 1 2 |\Omega_-\rangle \langle \Omega_-|
-
\frac 1 2 |\Omega_+\rangle \langle \Omega_-|
-
\frac 1 2 |\Omega_-\rangle \langle \Omega_+|
\label{eq:3490}\\
&=&
-
\frac 1 2
\frac{ 1  }{ \gamma_1 - \gamma_2 + \gamma_3} \rho_1
+
\frac 1 2
\frac{ \gamma_1 - \gamma_2 + 2 \gamma_3 }{\gamma_1 - \gamma_2 + \gamma_3 } \rho_2
+
\frac{1}{ \gamma_2 (\gamma_1 + \gamma_3 ) } \rho_3
-
\frac 1 2  \rho_4
-
\frac 1 2 \rho_7
\end{eqnarray}
leads, in exact resonance, to
\begin{eqnarray}
\rho(t)
&=&
-
\frac 1 2
\frac{ 1  }{ \gamma_1 - \gamma_2 + \gamma_3}
e^{- \frac{ \gamma_1 + \gamma_3 }{ 2 } t}
\rho_1
+
\frac 1 2
\frac{ \gamma_1 - \gamma_2 + 2 \gamma_3 }{\gamma_1 - \gamma_2 + \gamma_3 }
e^{- \frac{ \gamma_2 }{ 2 } t}
\rho_2
+
\frac{1}{ \gamma_2 (\gamma_1 + \gamma_3 ) } \rho_3
\nonumber
\\
&&
-
\frac 1 2
e^{- 2 i {q} t }
e^{- \frac{ \gamma_1 + \gamma_2 + \gamma_3 }{4} t}
\rho_4
-
\frac 1 2
e^{2 i {q} t }
e^{- \frac{ \gamma_1 + \gamma_2 + \gamma_3 }{4} t}
\rho_7,\\
p_g^{\rm irr}(t)
&=&
1
-
\frac 1 4
\frac{ \gamma_1 - \gamma_2 + 2\gamma_3 }{ \gamma_1 - \gamma_2 + \gamma_3 }
e^{- \frac{ \gamma_2 }{ 2 }  t}
-
\frac 1 4
\frac{ \gamma_1 - \gamma_2 }{ \gamma_1 - \gamma_2 + \gamma_3 }
e^{- \frac{ \gamma_1 + \gamma_3 }{ 2 } t}
-
\frac 1 2
e^{- \frac{ \gamma_1 + \gamma_2 + \gamma_3 }{4} t}
\cos 2  {q} t.
\label{eq:3500}
\end{eqnarray}
\section*{Appendix B: Quantum weak law of large numbers}

Consider some projector $P_k=|k\rangle\langle k|$ and a state vector $|\psi\rangle=\sum_k\psi_k |k\rangle$. The probability of observing a property represented by $P_k$ is
$|\psi_k|^2$ and the eigenvalues of $P_k$ are 0 (``failure") and 1 (``success"). Now let us consider a system of $N$ copies of $|\psi\rangle$, i.e.
\be
|\Psi\rangle
&=&
\underbrace{
|\psi\rangle
\otimes\dots\otimes
|\psi\rangle}_N.
\ee
The operator that counts the number of successes in $N$ independent measurements of $P_k$ is
\be
\hat N_k
&=&
P_k\otimes 1\otimes\dots\otimes 1+\dots+1\otimes\dots \otimes 1\otimes P_k
\ee
Denoting $P_1=P_k$, $P_0=1-P_k$, we obtain spectral representation of $\hat N_k$ in the form
\be
\hat N_k
&=&
\sum_{s=0}^N\sum_{A_1+\dots+A_N=s}s\,P_{A_1}\otimes\dots\otimes P_{A_N},\quad A_j=0,1.
\ee
Let $f:[0,1]\to \bm R$ be some function. Spectral theorem defines $f(\hat N_k/N)$ as
\be
f(\hat N_k/N)
&=&
\sum_{s=0}^N\sum_{A_1+\dots+A_N=s}f(s/N)\,P_{A_1}\otimes\dots\otimes P_{A_N}.
\ee
For $N\to\infty$ the average
\be
\langle\Psi|f(\hat N_k/N)|\Psi\rangle
&=&
\sum_{s=0}^Nf(s/N){N\choose s} \big(|\psi_k|^2\big)^s\big(1-|\psi_k|^2\big)^{N-s}
\ee
converges to $f(|\psi_k|^2)$ (by virtue of the law of large numbers for the binomial distribution).
The latter implies also that
\be
\lim_{N\to\infty}\parallel f(\hat N_k/N)|\Psi\rangle-f(|\psi_k|^2)|\Psi\rangle\parallel=0.
\ee
An analogous proof can be formulated for projectors $P_k$ projecting on higher dimensional subspaces, for example for
$P_k=1\otimes|k\rangle\langle k|$. Attempts of formulating a quantum analogue of the strong law of large numbers lead to infinite tensor products, $N=\infty$, and thus to nonseparable Hilbert spaces. Field quantization in terms of reducible representations of the harmonic oscillator Lie algebra involves only a finite $N$, even if the number of frequencies allowed by given boundary conditions is infinite.

\subsection*{Appendix C: Naive regularization versus relativistic covariance}

Let $a(k)$ be a scalar field annihilation operator that transforms covariantly under Lorentz transformations: $U(L)^\dag a(k) U(L)=a(L^{-1}k)$, with $U(L)^\dag U(L)=1$.
The standard commutator
$[a(k),a(k')^\dag]=Z\delta(k-k')1$, where $Z$ is a constant and $\delta(k-k')$ a relativistically invariant Dirac delta, is relativistically covariant. The proof is trivial: On one hand, $U(L)^\dag [a(k),a(k')^\dag]U(L)=[U(L)^\dag a(k)U(L),U(L)^\dag a(k')^\dag U(L)]=[a(L^{-1}k),a(L^{-1}k')^\dag]$.
On the other hand, $U(L)^\dag Z\delta(k-k')1\, U(L)=Z\delta(k-k')1=Z\delta(L^{-1}k-L^{-1}k')1 $ since delta is invariant by assumption.
The proof fails if $Z$ is replaced by a nontrivial function $Z(k)$, i.e. in naive regularizations.

However, since $U(L)^{\dag}|k\rangle=|L^{-1}k\rangle$, a replacement of $Z(k)$ by $I(k)=1\otimes|k\rangle\langle k|$, or by a similar object, removes the difficulty: $U(L)^\dag I(k) U(L)=I(L^{-1}k)$. Simultaneously, in all averages the weak law of large numbers effectively replaces $I(k)$ by $Z(k)$ (see Appendix B). Another conclusion is that, as opposed to naive regularization, spectra of Hamiltonians do not depend on $Z(k)$, but on relativistically invariant eigenvalues $s/N$ (think of the dressed states we have described in the present paper).
This is why, following Finkelstein \cite{Finkelstein}, we speak of {\it regularization by quantization\/}.
For more details on relativistic aspects see e.g. \cite{MCKW}.


\begin{thebibliography}{99}
\bibitem{no jumps}This follows from the form (\ref{CCR3}) of $\uu N_\omega$ which, for $N>1$, should not be identfied with $\uu{\tilde N}_\omega
=\uu a_\omega^\dag\uu a_\omega$. Had we included the term $\uu{\tilde N}_\omega B$, we would have generated also the jumps involving $|\Omega_k(s)\rangle$. Note, however, that $[\uu a_\omega,\uu{\tilde N}_\omega]=\uu I_\omega\uu a_\omega=a_0(\omega)a_-(\omega)$, so the algebra involving $\uu{\tilde N}_\omega$ is not a Lie algebra.
\bibitem{eps=0}Unfortunately, this formula cannot be compared directly with the results from \cite{pra}, since we did not show explicitly the $T=0$ case (see our remarks under Eq.~(62) in \cite{pra}). The $T=0$ case is computed here in the Appendix.
\end{thebibliography}
\end{document}